\documentclass[12pt,preprint]{aastex}

\slugcomment{}
\shorttitle{F-Supergiant Coronae}
\shortauthors{T.\ R.\ Ayres}
\begin{document}

\title{Cracking the Conundrum of F-Supergiant Coronae}

\author{Thomas R.\ Ayres}

\affil{Center for Astrophysics and Space Astronomy,\\
389~UCB, University of Colorado,
Boulder, CO 80309;\\ Thomas.Ayres@Colorado.edu}

\begin{abstract}
{\em Chandra}\/ X-ray and {\em HST}\/ far-ultraviolet (FUV) observations of three early-F supergiants have shed new light on a previous puzzle involving a prominent member of the class: $\alpha$~Persei (HD\,20902: F5~Ib).  The warm supergiant is a moderately strong, hard coronal ($T\sim 10^7$~K) X-ray source, but has ten times weaker ``sub-coronal'' \ion{Si}{4} 1393~\AA\ ($T\sim 8\times10^4$~K) emissions than early-G supergiants of similar high-energy properties.  The $\alpha$~Per X-ray excess speculatively was ascribed to a close-in hyperactive G-dwarf companion, which could have escaped previous notice, lost in the glare of the bright star.  However, a subsequent dedicated multi-wavelength imaging campaign failed to find any evidence for a resolved secondary.  The origin of the $\alpha$~Per high-energy dichotomy then devolved to: (1) an {\em unresolved}\/ companion; or (2) intrinsic coronal behavior.  Exploring the second possibility, the present program has found that early-F supergiants do appear to belong to a distinct coronal class, characterized by elevated X-ray/FUV ratios, although sharing some similarities with Cepheid variables in their transitory X-ray ``high states.''  Remarkably, the early-F supergiants now are seen to align with the low-activity end of the X-ray/FUV sequence defined by late-type dwarfs, suggesting that the disjoint behavior relative to the G supergiants might be attributed to thinner outer atmospheres on the F types, as in dwarfs, but in this case perhaps caused by a weakened ``ionization valve'' effect due to overly warm photospheres.
\end{abstract}


\section{INTRODUCTION}

A 2010 {\em Hubble}\/ Cosmic Origins Spectrograph (COS) SNAPshot (partial orbit fillers in the {\em HST}\/ schedule) of the X-ray bright, mid-F supergiant $\alpha$~Persei found unexpectedly weak far-ultraviolet (FUV) emissions of ``hot lines'' like \ion{Si}{4} 1393~\AA, at least compared to somewhat cooler early-G supergiants of similar X-ray luminosity (Ayres 2011).  The only other well-observed F supergiant at the time, Canopus ($\alpha$~Car; HD\,45348; F0\,Ib), also displayed an unusual X-ray/FUV excess.  Initial speculation was that the $\alpha$~Per X-ray anomaly could be explained, most simply, by a previously unseen G-dwarf companion, partly motivated by a small, but suspicious, offset ($\sim 9\arcsec$) of the $\alpha$~Per X-ray source in a 1993 {\em ROSAT}\/ pointing.  The intermediate-mass ($\sim 7\,M_{\odot}$) supergiant must be very young, $\sim 50$~Myr (based on isochrone fitting of the eponymous cluster: e.g., Makarov 2006), consequently a coeval late-type dwarf could be extremely X-ray-bright owing to the strong association of high coronal activity with fast rotation, and thus also youth, among low-mass Main sequence stars (e.g., Pallavicini et al.\ 1981).  Further, a close-in dwarf companion, with $\Delta{V}\sim +9$, easily could have escaped notice behind the glare of the bright supergiant.

Thus ensued a multi-wavelength campaign to search for a spatially resolved companion.  The program extended from the ground, with visible-light coronography at Apache Peak Observatory; to the near-UV, with {\em HST}\/ Wide Field Camera 3 (WFC3); and X-rays, with {\em Chandra}\/ High Resolution Camera (HRC-I) (Ayres 2017 [A17]).  The campaign also collected additional FUV spectra of the F supergiant with COS; especially below 1300~\AA, forbidden to the original SNAP owing to concerns that the (then unknown) \ion{H}{1} 1215~\AA\ Ly$\alpha$ emission might violate COS detector bright limits.  The new program, however, failed to recover a resolved companion, even well inside the already small separation suggested by {\em ROSAT.}\/  Further,  FUV ``flux-flux'' diagrams, specifically comparing hot \ion{Si}{4} versus cooler chromospheric semi-permitted \ion{O}{1}] (1355~\AA: $T< 10^4$~K), placed $\alpha$~Per (and Canopus) plausibly on a low-activity extension of the power-law trend defined by the G supergiants: the weak sub-coronal emission of the two F supergiants was paralleled by weak chromospheric oxygen emission.  The possible sub-coronal/chromospheric continuity with the cooler supergiants reinforced the idea that the X-ray behavior of the warmer stars was the main difference.  

Given that the X-ray luminosity of $\alpha$~Per was exactly right for a young G dwarf, the joint imaging campaign still did not eliminate the possibility that the high-energy anomaly was due to an {\em unresolved}\/ secondary star; but there was an equally viable alternative, namely that the odd behavior was intrinsic to the F supergiant class itself.  Perhaps significantly, the two Cepheid variables with phase-diverse high-energy measurements at the time -- $\delta$~Cephei (F5~Iab) and $\beta$~Doradus (F8/G0~Ib) (Engle et at.\ 2014; Engle 2015; Engle et at.\ 2017) -- had FUV ``high states'' that roughly connected to the early-G supergiants in an X-ray/\ion{Si}{4} flux-flux diagram, but X-ray high states that pointed more toward the region occupied by $\alpha$~Per and Canopus.  (The Cepheid FUV high states occur during pulsational phases when the X-rays are low, and vice versa; and both high states are transitory [ibid].)

The obstinate ``F-supergiant conundrum'' -- extrinsic interlopers versus intrinsic behavior --  inspired a new joint {\em Chandra/HST}\/ program, this time to explore the second alternative, by probing the luminous early-F stars more broadly.  To preview the results, the ``anomalous'' X-ray behavior of the early-F supergiants appears to be, in fact, perfectly normal for the class, and quite disjoint from their cooler G-type cousins only a little further along the 5--10~$M_{\odot}$ evolutionary tracks.  In fact, the F supergiants seem, counterintuitively, to follow the X-ray/FUV flux-flux sequence defined by cool dwarfs, albeit at the extreme low end, in spite of the ostensibly vast gulf between their physical properties.  

The paper is organized as follows:  \S{2} describes the target stars, the {\em HST}\/  and {\em Chandra}\/ observations, and associated measurements; \S{3} offers various comparisons between the F-supergiant observables and those of the cooler G-type supergiants as well as representative G--M Main sequence stars; and \S{4} weighs these relationships with regard to a new F-supergiant conundrum, namely their very un-supergiant-like behavior in chromospheric and sub-coronal tracers.

\begin{figure}[ht]
\figurenum{1}
\vskip  0mm
\hskip  +10mm
\includegraphics[width=0.8\linewidth]{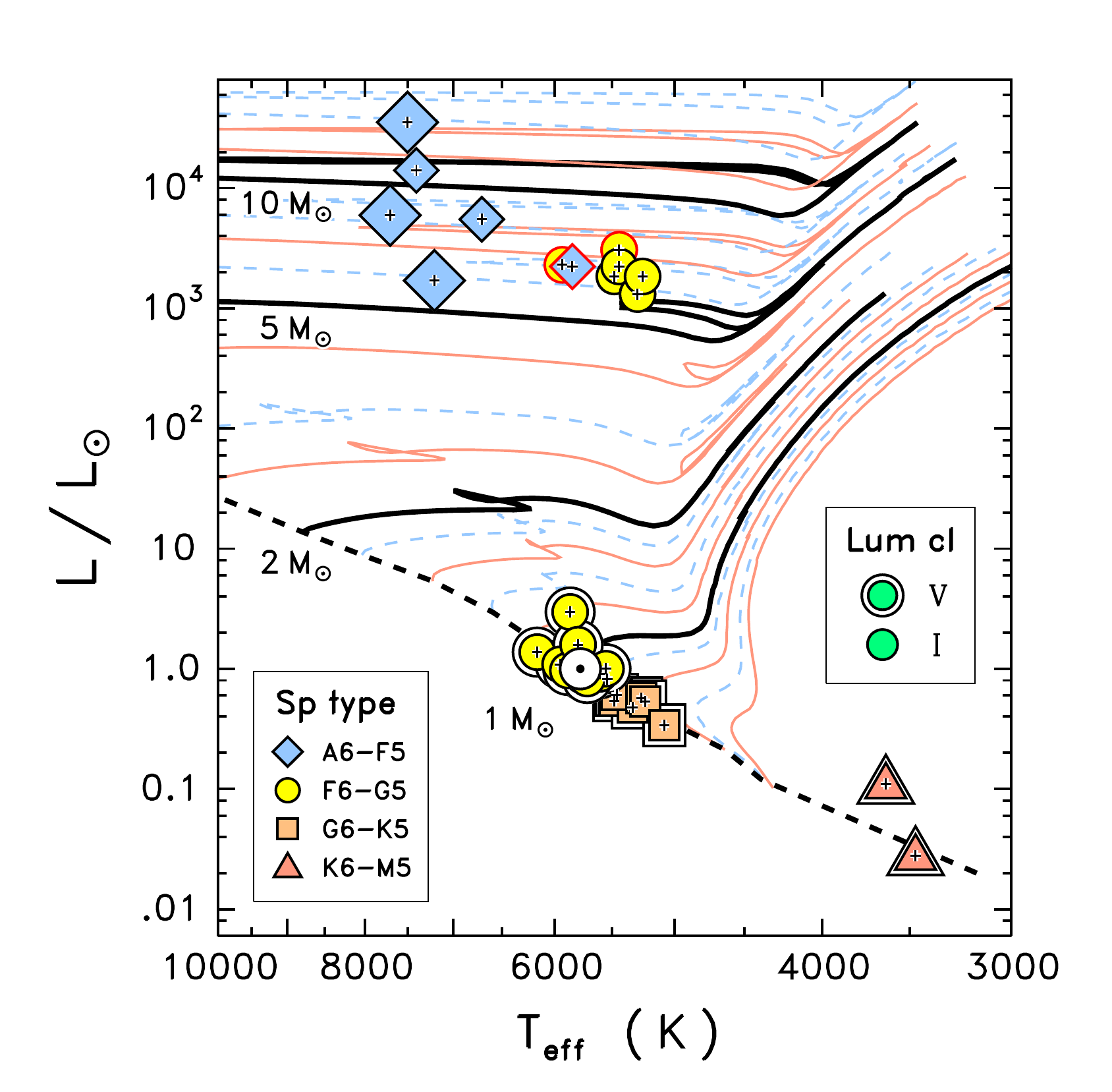} 
\caption[]{\small
H--R diagram of representative dwarfs and supergiants, subjects of various comparisons described later.  Stellar types are marked by symbols and colors according to the two keys.  Location of the Sun is marked $\odot$.  Thick and thin solid, and light-dashed, curves are Padova evolutionary tracks (purely illustrative).  The diagonal thick dashed curve is the Zero Age Main Sequence (ZAMS).  Selected tracks (dark, thicker curves) are marked with the ZAMS masses.  Red-outlined symbols (upper middle) are Classical Cepheids.  Main subjects of the present study are the three larger blue diamonds in the upper left hand corner of the diagram; those of the previous study, $\alpha$~Per and Canopus, are the smaller blue (non-Cepheid) diamonds.  
}
\end{figure}

\clearpage
\section{OBSERVATIONS}

\subsection{Target Stars}

Properties of the three selected F supergiant targets, plus the previous objects of interest $\alpha$~Per and Canopus, are summarized in Table~1.  The new targets were chosen from a larger sample of visually bright members of the early-F class, which had existing but inconclusive evidence of X-ray emission, and which were found to have negligible interstellar extinction by Bersier (1996).  (Low reddening is important mainly for FUV sensitivity.)  Derived luminosities place the new F supergiants, as well as $\alpha$~Per and Canopus, in the mass range 5--10~$M_{\odot}$.   Figure~1 is an H--R diagram comparing the new targets to the F and G supergiants and other reference stars of A17.  (Evolutionary tracks in the figure are from http://stev.oapd.inaf.it/cgi-bin/cmd).

Key parameters of the F supergiants were abstracted mostly from SIMBAD, although effective temperatures were based on the PASTEL Catalog of Soubrain et al.\ (2016), averaging multiple entries, if necessary (see A17).  The consensus values are in general agreement with Flower's (1996) tables of $T_{\rm eff}$ versus (de-reddened) $(B-V)$ colors.  Also, ranges are provided for a few parameters, such as spectral classifications, where indicated by SIMBAD or otherwise.  Of the three new targets, $\theta$~Sco is an unusually fast rotator for a supergiant ($\upsilon\sin{i}\sim 105$~km s$^{-1}$: Snow et al.\ 1994); the others are more normal ($\upsilon\sin{i}\lesssim 20$~km s$^{-1}$).  Bolometric corrections were taken from Flower's tables, for the de-reddened $(B-V)$ colors.

Although the new targets, and $\alpha$~Per and Canopus, all were noted in Bersier (1996) as having essentially zero reddening, the catalog of Snow et al.\ (1994) offers an alternative view.  For example, in the main table of Bersier (Table~2: ``computed color excesses''), $\alpha$~Per is assigned $E(B-V)= 0.01$; while in Bersier's Table~1 (``stars used for the calibration'') it is listed as $E(B-V)= 0.04$; but in Snow et al.\ (1994), the color excess is much larger, 0.15 magnitudes.  The enhanced extinction would affect estimates of effective temperature from $(B-V)$, as well as reddening corrections for X-ray and FUV fluxes.  The Snow catalog also cites a larger reddening for $\theta$~Sco (0.13 magnitudes), although the Snow values for $\iota$~Car, $\alpha$~Lep, and Canopus are about as small as the Bersier limits.  In the present study, the Snow color excesses were adopted (second values in Table~1, here), since they were derived from more diverse, multi-spectral evidence than the Bersier study (based solely on Geneva photometry).  Incidentally, the Snow effective temperatures for $\alpha$~Car and $\theta$~Sco are very close to the consensus values adopted here, while $\alpha$~Per, $\alpha$~Lep, and $\iota$~Car are systematically higher, but only by about 200~K.

T.~J.~J.~See (1896) reported that $\theta$~Sco has a faint ($\sim13$th magnitude) companion with a separation of about 6$\arcsec$.  The {\em Hipparcos}\/ catalog lists a companion at about the same position angle and separation found by See, but with a differential brightness of only 3.4 magnitudes ($\Delta{H_{\rm p}}$).  The Notes in the Washington Visual Double Star Catalog (Mason et al.\ 2001, as updated in the VizieR On-Line Data Catalogs) suggest that the {\em Hipparcos}\/ binary solution might be in error.  Given the tight separation, the putative companion almost certainly would be physically associated -- and coeval -- with the supergiant, and thus be an un-evolved Main sequence star, around spectral type late-B given the {\em Hipparcos}\/ brightness deficit.  However, the {\em FUSE}\/ spectrum of $\theta$~Sco, taken through the $30{\arcsec}{\times}30{\arcsec}$ LWRS aperture (which would include $\theta$~Sco ``B''), displays only a faint continuum in the sub-Ly$\alpha$ wavelength region ($\lambda< 1200$~\AA: $f_{\lambda}< 10^{-14}$ erg cm$^{-2}$ s$^{-1}$ {\AA}$^{-1}$), consistent with the new {\em HST}\/ COS 2.5$\arcsec$-diameter Primary Science Aperture (PSA) observation described later (which would exclude a 6$\arcsec$ companion).  Sub-Ly$\alpha$ flux densities for a late-B MS star would be several orders of magnitude larger.  The much fainter secondary reported by See ($\Delta{V}\sim +11$) would correspond to a late-K dwarf at the distance of $\theta$~Sco; likely coronally active, if coeval with the young primary.  However, the {\em Chandra}\/ HRC-I image of $\theta$~Sco, described later, shows no evidence for additional counts at the 6$\arcsec$ separation and PA reported by See (and {\em Hipparcos}), or anywhere close to the (already barely detected) central source for that matter (the nearest bright X-ray object is half an arcminute away).

\begin{deluxetable}{cccccccccc}
\tabletypesize{\footnotesize}
\tablenum{1}
\tablecaption{F Supergiant Targets and Stellar Parameters}
\tablecolumns{10}
\tablewidth{0pt}
\tablehead{\colhead{Name} & \colhead{HD No.} & \colhead{Type} & 
\colhead{$V$}  & \colhead{$(B-V)$}  &  \colhead{$E(B-V)$}  &  \colhead{B.C.} &
 \colhead{$d$} &  \colhead{$T_{\rm eff}$} &  \colhead{$\upsilon\sin{i}$}\\
 &  &  & \multicolumn{4}{c}{-----~~~~~~~(magnitudes)~~~~~~~-----} & 
 \colhead{(pc)} & \colhead{(K)}  & \colhead{(km s$^{-1}$)} \\
\colhead{(1)} & \colhead{(2)} & \colhead{(3)} & \colhead{(4)} & \colhead{(5)}  & 
\colhead{(6)}  & \colhead{(7)} & \colhead{(8)} & \colhead{(9)} & \colhead{(10)}
} 
\startdata
$\alpha$~Persei    &  20902  &               F5\,Ib &  $+1.79$ &  $+0.48$ &  $0.04$--$0.15$ &  $+0.02$ &  155   & 6700 &  $<18$\\ 
$\alpha$~Leporis    &  36673  &                 F0Ib &  $+2.57$ &  $+0.20$ &  $0.00$--$0.02$ &  $+0.03$ &  680  & 7500  &  13--21 \\                                        
$\alpha$~Carinae    &   45348  &  A9\,II--F0\,Iab &  $-0.74$ &  $+0.15$ &  $0.00$--$0.00$ &  $+0.03$ &    95  & 7400 & 9 \\   
$\iota$~Carinae       &   80404  &  A7\,Ib--F0\,Ib &  $+2.26$ &  $+0.18$ &  $0.00$--$0.04$ &  $+0.03$ &  235  & 7700 & 10 \\                                                        
$\theta$~Scorpii     & 159532  &  F1\,III--F0\,Ib &  $+1.86$ &  $+0.40$ &  $0.00$--$0.13$ &  $+0.03$ &   92   & 7200 & 105 \\                        
\enddata
\end{deluxetable}

\begin{deluxetable}{ccrccc}
\tabletypesize{\small}
\tablenum{2}
\tablecaption{{\em HST}\/ COS and {\em Chandra}\/ HRC-I Observing Logs}
\tablecolumns{6}
\tablewidth{0pt}
\tablehead{\colhead{Dataset} & \colhead{UT Start} & \colhead{$t_{\rm exp}$} & 
\colhead{Splits}  & \colhead{Aperture}  &  \colhead{Grating} \\
\colhead{(1)} & \colhead{(2)} & \colhead{(3)} & \colhead{(4)} & \colhead{(5)}  & \colhead{(6)}  
} 
\startdata
\cutinhead{HD\,36673\,=\,$\alpha$~Lep}
ldc0a0010    &  2017-04-17.805  &  1672  &  1--4   &     PSA      &  G130M--1327 \\
ldc0a0020    &  2017-04-17.861  &  2589  &  1--4   &     PSA      &  G130M--1291 \\
ObsID\,18912       & 2017-04-20.163   &  19050 &  \nodata   &    $30\arcmin{\times}30\arcmin$  &   \nodata \\ 
\cutinhead{HD\,80404\,=\,$\iota$~Car}
ldc0b0010    &  2017-04-04.961  &  1936  &  1--4   &     PSA      &  G130M--1327 \\
ldc0b0020    &  2017-04-05.025  &  2825  &  1--4   &     PSA      &  G130M--1291 \\
ObsID\,18913        & 2017-06-11.899   &  18870 &  \nodata    &    $30\arcmin{\times}30\arcmin$  &   \nodata  \\ 
\cutinhead{HD\,159532\,=\,$\theta$~Sco}
ldc0c0010    &  2017-03-19.830  &  1804  &  1--4   &     PSA      &  G130M--1327 \\
ldc0c0020    &  2017-03-19.894  &  2677  &  1--4   &     PSA      &  G130M--1291 \\
ObsID\,18914        & 2017-03-23.662   &  19770 &  \nodata    &    $30\arcmin{\times}30\arcmin$  &   \nodata  \\ 
\enddata
\tablecomments{Col.~1 prefix  ``ldc0'' refers to {\em HST}\/ COS (program GO-14848); ``ObsID'' to {\em Chandra}\/ HRC-I (program 18200144).  Col.~3 is total exposure (in seconds), corrected for dead time for HRC-I.  Col.~4 ``Splits'' are the four standard FP-POS steps for COS, with duration $t_{\rm exp}/4$.  Col.~5 ``Aperture'' is $2.5\arcsec$-diameter Primary Science Aperture (PSA) for COS, and field-of-view for HRC-I.  Col.~6 lists grating--$\lambda_{\rm cen}$(\AA) for COS.}
\end{deluxetable}

\begin{figure}[ht]
\figurenum{2}
\vskip   0mm
\hskip   -3mm
\includegraphics[width=\linewidth]{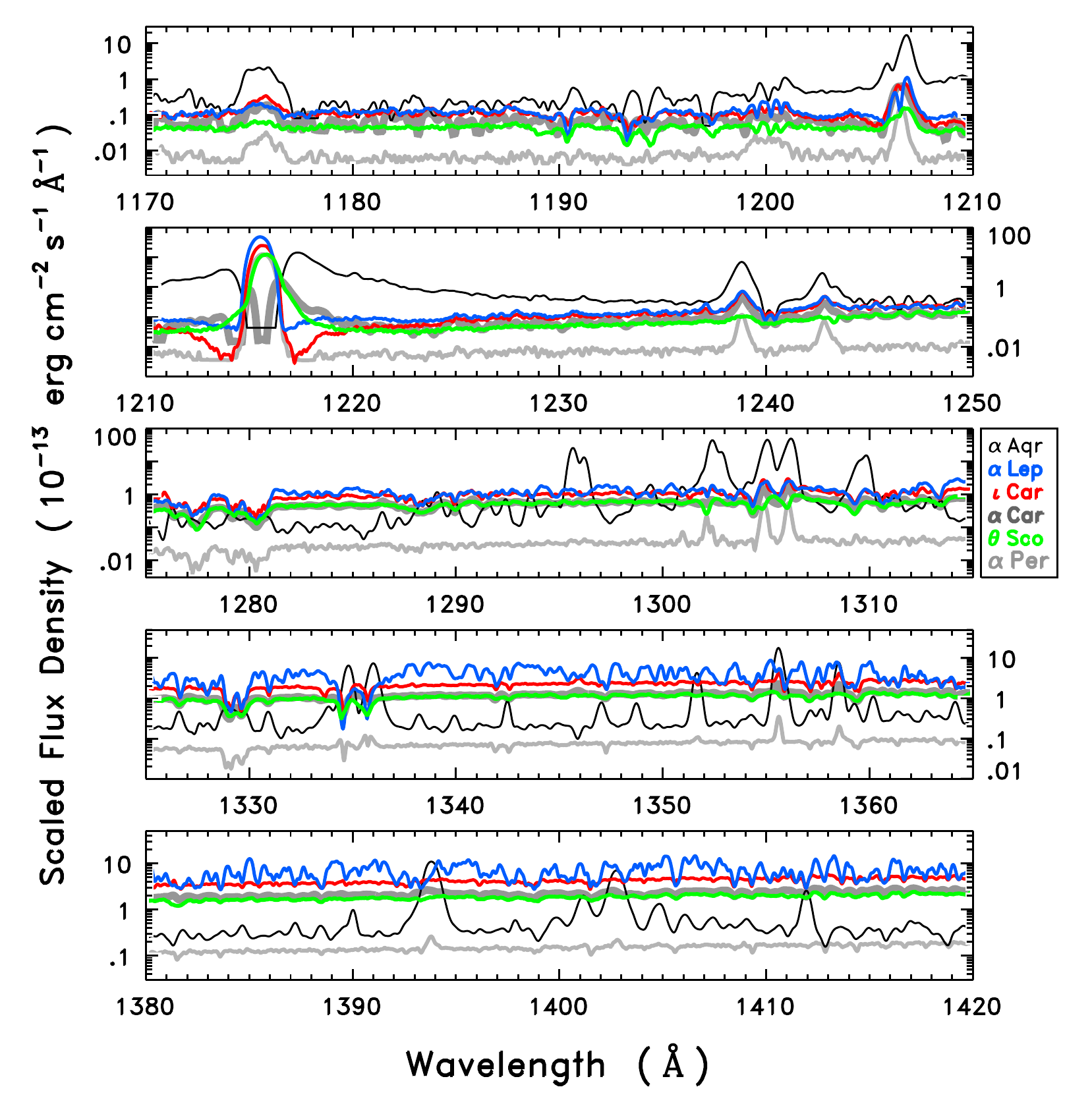} 
\vskip  0mm
\caption[]{\small
Gaussian smoothed (30~km s$^{-1}$ FWHM) FUV spectra of the three new F supergiants (thinner curves, various colors) compared with reference F stars $\alpha$~Per (thin lighter gray curve) and Canopus (thick darker gray curve) and G supergiant $\alpha$~Aqr (G2~Ib: thin black curve).   (Legend at the right hand side of the diagram shows specific order of the tracings in the middle panel.)  The flux density scale refers to $\alpha$~Per; note the many decades covered.  The other supergiants were adjusted according to their $V$-band intensities relative to $\alpha$~Per (equivalent to a bolometric luminosity normalization).  The Ly$\alpha$ cores of the four COS stars ($\alpha$~Per and the three new F supergiants) are dominated by hydrogen geocoronal emission through the 2.5$\arcsec$-diameter PSA, whereas the two STIS objects ($\alpha$~Car and $\alpha$~Aqr) show Ly$\alpha$ ISM absorption.
}
\end{figure}

\clearpage
\subsection{{\em HST}\/ COS FUV Spectroscopy}

{\em IUE}\/ low-resolution FUV spectra of the three new early-F supergiants showed normal energy distributions for the class: bright at 1500~\AA, but fading rapidly toward shorter wavelengths.  That, and the unexpectedly faint sub-coronal emission lines of $\alpha$~Per, favored {\em HST's}\/ Cosmic Origins Spectrograph (COS) over higher resolution, but less sensitive, Space Telescope Imaging Spectrograph (STIS) (which nevertheless had been used successfully for UV-brighter Canopus).  

The COS program was carried out in the 2017 March/April time frame, in three visits (one per target) of two orbits each, as summarized in Table~2.   The spectral range of interest was the short-FUV (1150--1450~\AA), captured by COS grating G130M, hosting a variety of key plasma diagnostics (including Ly$\alpha$).  Extending the program to longer wavelengths was judged futile, because the rapidly rising F-star continuum quickly would obliterate any of the normally expected hot lines (e.g., \ion{C}{4} 1548~\AA).  In fact, even \ion{Si}{4} 1393~\AA\ -- clearly detected in $\alpha$~Per (although weak) -- was not visible in any of the new, hotter F supergiants.  The resolving power ($\lambda/\Delta\lambda$) of G130M is about 18,000 (17~km s$^{-1}$), fully adequate except perhaps for the narrowest interstellar features.

At the beginning of each visit, the target was acquired in dispersed FUV light using G130M itself.  The 1327~\AA\ CENWAVE was chosen for the purpose, because it is the reddest of the G130M settings, and thus collects the largest flux from the bright F-star continuum, minimizing exposure times for the multiple short acquisition pointings.  Initially, a 9-step raster search located the target coarsely.  The centering then was refined with a PEAKXD (``peak-up'' [centroiding] in the cross-dispersion direction), followed by a PEAKD (peak-up along the dispersion direction).  The remainder of the first orbit was filled by four equal-duration G130M-1327 exposures, 400--500~s depending on the target visibility; at the (four) standard FP-POS steps (small grating rotations intended to mitigate fixed-pattern noise).  The minimal Ly$\alpha$ emission of $\alpha$~Per, from the deep COS pointings of the previous program, allayed any concerns over a bright limit violation for the new targets, which allowed detector side B ($\lambda< 1300$~\AA) to be activated.   The second orbit featured a similar sequence of G130M exposures, this time with CENWAVE 1291~\AA, also at the four standard FP-POS steps.  Exposure times for the equal-duration ``splits'' were 650--700~s, again depending on the target visibility.  The two distinct grating settings eliminated the (detector) spectral gap otherwise present if only a single CENWAVE had been used.

The two sets of G130M exposures from each visit were processed through the {\sf CALCOS} pipeline, which combined the four separate FP-POS sub-exposures of each CENWAVE.  The co-addition suppressed pipeline-flagged fixed-pattern defects such as grid wire shadows.  The two independent FUV tracings then were spliced together, to eliminate the detector gaps.  In addition, spatial/spectral maps of the two CENWAVEs were assembled directly from the event lists (see, e.g., Ayres 2015).  There was no evidence for any cross-dispersion asymmetries or spectrum doubling that might indicate the presence of a partially resolved companion.  Time series of key features, such as \ion{Si}{3} 1206~\AA, \ion{N}{5} 1238~\AA, and \ion{O}{1} 1304~\AA\ + 1306~\AA, also were extracted from the event lists (ibid).  No obvious transient behavior (such as flaring) was seen in any of the targets, other than modest changes in the \ion{O}{1} resonance lines owing to variable atomic oxygen skyglow.

The merged COS G130M FUV spectra of the three new F supergiants are compared in Figure~2 to the deep COS spectrum of $\alpha$~Per and the high-quality STIS tracings of brighter Canopus and representative G-supergiant $\alpha$~Aqr (G2~Ib).  Each spectrum was smoothed by a Gaussian filter with a FWHM of 2 COS resolution elements (resel).  The $\alpha$~Per flux densities are displayed as observed; the other supergiants were adjusted according to their relative visual fluxes (${\times}10^{+(V_{\star}-V_{\alpha\,{\rm Per}})/2.5}$).  The spectra were not corrected for reddening.  

All the F supergiants display continuum-dominated energy distributions, with conspicuous absorption structure longward of $\sim$1310~\AA.  The three new objects cluster around the Canopus spectrum in relative flux density, significantly elevated compared to somewhat cooler $\alpha$~Per.  Nevertheless, yellow supergiant $\alpha$~Aqr also displays an enhanced continuum, even above that of $\alpha$~Per on the $V$--adjusted scale, despite the significantly cooler ($\sim 1000$~K) photosphere of the G star.  The outer atmosphere of $\alpha$~Aqr is optically thick enough that the FUV continuum has a significant contribution from the hotter chromosphere (exponentially Planck-weighted relative to the lower temperature photosphere).  At the same time, the G supergiant clearly is emission-line-dominated compared to the warmer F-type supergiants.  Particularly striking are the prominent \ion{Si}{4} 1393~\AA\ + 1402~\AA\ features of the G supergiant, while only those of $\alpha$~Per -- but not the other F supergiants -- are visible (and then only barely so).

Note, also, that all the F supergiants have strong ground-configuration \ion{C}{1} absorptions (e.g., 1276--1280~\AA, 1329--1330~\AA), whereas the same multiplets are in emission in the G supergiant.  The absorption behavior in the F stars indicates formation within a photosphere, namely temperatures falling outward with increasing altitude, so that the optically thicker carbon line cores arise at lower temperatures and thus lower intensities than the thinner continuum, which is emitted from deeper, hotter layers.  Conversely, the reversal from absorption to emission in the G supergiant points to formation in an atmosphere with temperatures rising outward, namely a chromosphere.  Even the higher temperature \ion{C}{2} 1335~\AA\ multiplet ($T\sim 2{\times}10^4$~K in ionization equilibrium), is in absorption in all the new F supergiants, although the extent to which interstellar or circumstellar absorption contributes to the apparent features is unknown.

On the other hand, the likely low opacity (high lower-level excitation energy) \ion{C}{3} multiplet at 1175~\AA\  ($6\times10^4$~K), is in emission in all the objects, at about the same relative strength in the three new F supergiants, similar to Canopus, but stronger than in $\alpha$~Per.  So too are the \ion{Si}{3} 1206~\AA\ features ($6\times10^4$~K), although the cores are affected to varying degrees by central absorptions, partly interstellar but possibly also circumstellar in some cases.  The similarity among the three new F supergiants and Canopus extends out to the key \ion{N}{5} doublet (1238~\AA\ + 1242~\AA), hottest ($T\sim 2\times10^5$~K) of the sub-coronal resonance transitions in the FUV.  Further, the \ion{O}{1} resonance triplet near 1305~\AA\ is in emission in all the F supergiants (less obvious in $\theta$~Sco in this view), and extremely prominent in G supergiant $\alpha$~Aqr; although in all cases partly obscured by sharp ISM absorptions in the 1302~\AA\ resonance line, and circumstellar absorptions, especially in ``subordinate'' 1304~\AA\ and 1306~\AA, in several of the stars ($\alpha$~Car and the three new F supergiants, as well as $\alpha$~Aqr).  The oxygen intersystem transition at 1355~\AA, which played an important role in the preceding study of $\alpha$~Per, is seen clearly in emission in only one of the new targets, $\iota$~Car.  In the others, it is strongly affected by neighboring \ion{C}{1} 1355~\AA\ absorption in the red wing of the \ion{O}{1}] feature. 

One conspicuous difference among the stars is \ion{H}{1} Ly$\alpha$.  It is absent in $\alpha$~Per and $\alpha$~Lep (to the extent that can be judged given the strong geocoronal contamination); relatively narrow, and red-asymmetric (stronger red peak) in both Canopus and $\theta$~Sco; with deep absorption wings in $\iota$~Car (and possibly weaker absorption wings in $\alpha$~Lep); but is very broad and strong -- although still red-asymmetric -- in $\alpha$~Aqr.  The Canopus Ly$\alpha$ profile, from STIS (almost geocorona-free), displays a distinct, sharp-edged absorption cut-out blueward of the presumably mostly interstellar absorption core, which likely indicates a current expanding wind or an archaic shell of material from a prior evolutionary stage (Brown et al.\ 2003).  The $\theta$~Sco profile from COS, outside the geocoronal dominated core, is structurally similar to that of $\alpha$~Car, in width and asymmetry, so the same assessment applies.

A closer view of selected spectral regions is provided in Figures~3a--3c, now on a linear scale.  The flux densities are as recorded by COS, without any correction for reddening, smoothed by 1 resel and shifted into the stellar frame according to the photospheric radial velocity.  No effort was taken to adjust the wavelength scales for the small distortions known to affect them (e.g., Ayres 2015), which are insignificant for the integrated fluxes illustrated in the figures.  Details of the measurement strategy, and particulars for each spectral feature, can be found in A17.  Results are summarized in Table~3.   The normally bright C~{\footnotesize II} 1335~\AA\ emission multiplet was purely in absorption in all the new stars, and thus not reported here.

No compensation was made for the possible circumstellar absorptions affecting the \ion{Si}{3} and \ion{O}{1} resonance lines in several of the stars.  Whereas an interstellar absorption would simply subtract flux from the intrinsic profile, the case of an expanding circumstellar shell or wind is more subtle.  Depending on the optical thickness and geometry of the flow, photons scattered out of the blueward absorption feature, formed on the front side of the star, could be compensated by photons at other frequencies, especially in the red peak, scattered into the line of sight from the other side of the flow on the back side of the star (``P-Cygni'' effect).  Further, the intrinsic shapes of the resonance lines are not known, and making the assumption that they were, say, Gaussian, might overestimate a correction, especially if the features possessed an intrinsic central reversal due to high opacity.  In any event, the maximum error in not compensating for the ``missing'' flux, if indeed it truly is missing, would be factors of 2, or so; not significant for the (logarithmic) flux-flux comparisons illustrated later.  For the specific key comparison \ion{Si}{3} versus \ion{O}{1}, the lack of missing-flux compensation would be less important, because both resonance transitions could be nearly equally affected.

\begin{figure}[ht]
\figurenum{3a}
\vskip  0mm
\hskip  -5mm
\includegraphics[width=\linewidth]{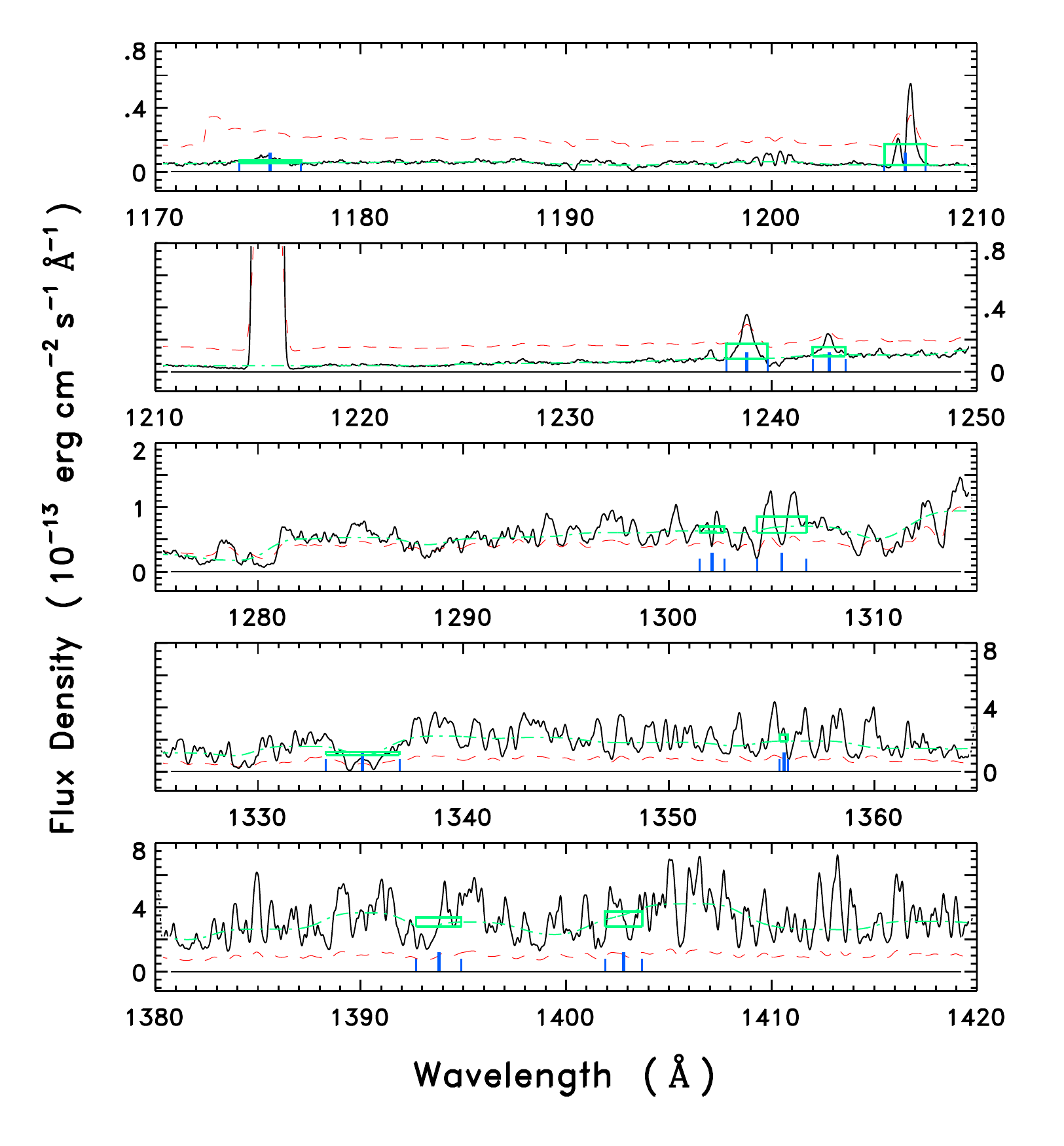} 
\vskip  -10mm
\caption[]{\small
Medium-resolution {\em HST}/COS FUV spectra of F supergiant  $\alpha$~Lep.  The thin dark curve is the observed spectrum, smoothed by 1 resel; red dashed curve is $50\times$ the photometric error (per resel).  The thin green dot-dashed curve is an estimated continuum level based on long-range filtering of the fluxes.  (The continuum modeling is optimized for emission lines on top of a structured background, and does poorly if the spectrum is absorption-line dominated, as for the longer wavelength segments of the F supergiants.)  Vertical blue ticks highlight the integration bandpasses for the specific single features, or multiplets, of interest.  Green boxes schematically illustrate the inferred integrated fluxes.  (In the flux integrations, the continuum was assumed to be constant over the extraction bandpass; the intensity value [lower edge of green box] was taken as the minimum of the continuum distribution evaluated in a slightly wider region.)
}
\end{figure}

\begin{figure}[ht]
\figurenum{3b}
\vskip  0mm
\hskip  -5mm
\includegraphics[width=\linewidth]{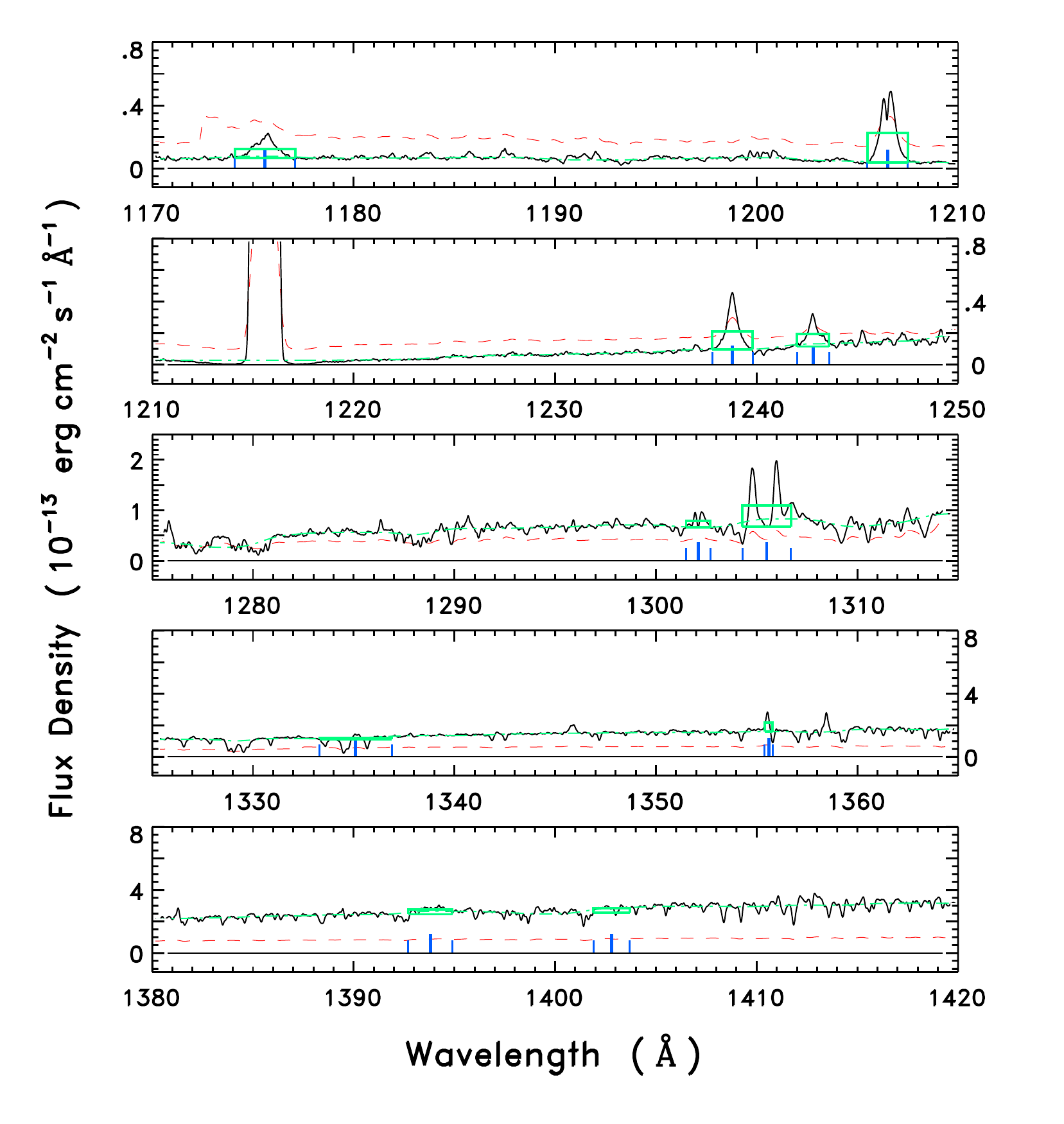} 
\vskip  0mm
\caption[]{\small
Same as Fig.~3a, for $\iota$~Car.
}
\end{figure}

\begin{figure}[ht]
\figurenum{3c}
\vskip  0mm
\hskip  -5mm
\includegraphics[width=\linewidth]{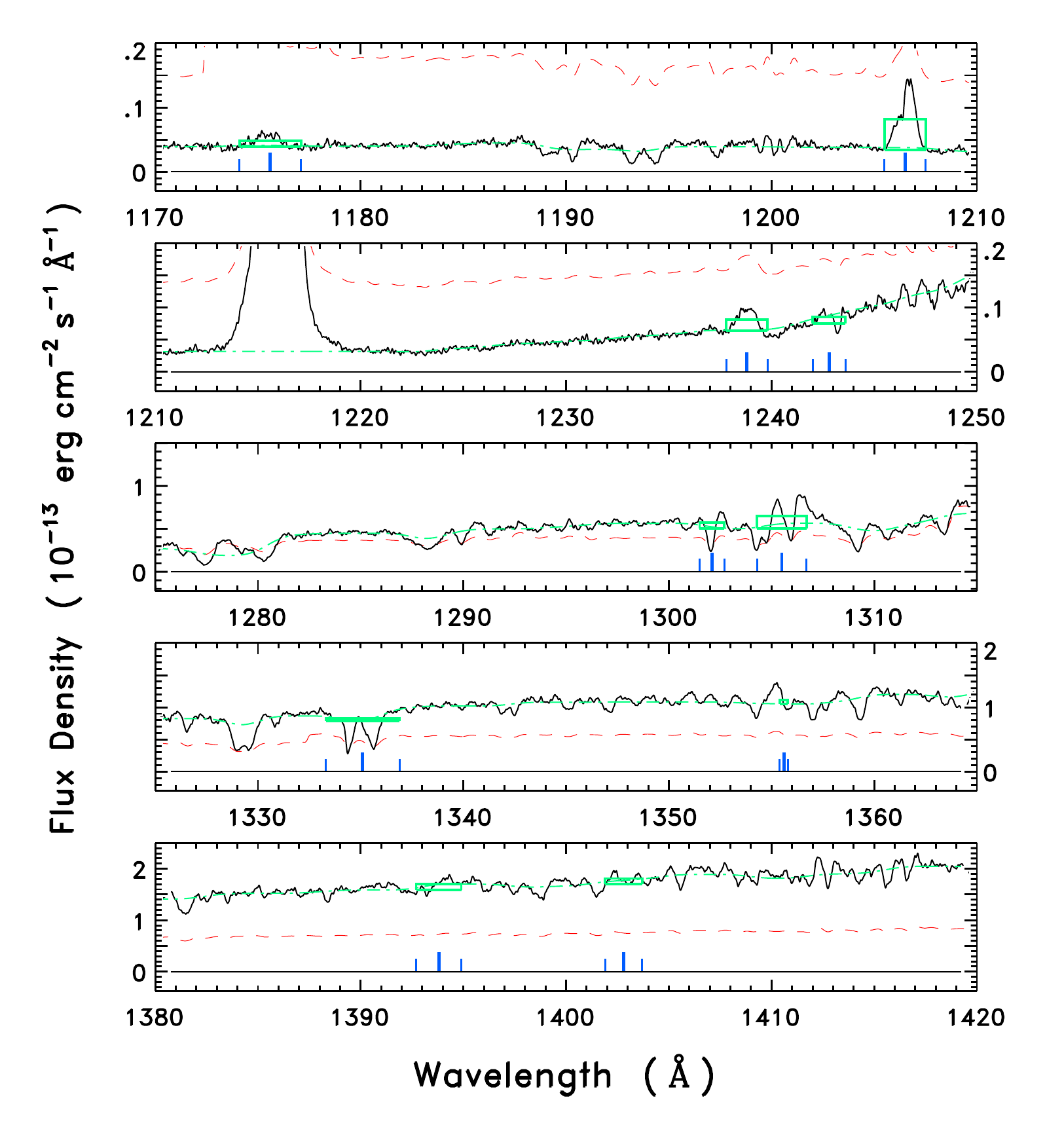} 
\vskip  0mm
\caption[]{\small
Same as Fig.~3a, for $\theta$~Sco.
}
\end{figure}

\clearpage
\begin{deluxetable}{ccccc}
\tablenum{3}
\tablecaption{Far-Ultraviolet Fluxes from {\em HST}\/ COS}
\tablecolumns{5}
\tablewidth{0pt}
\tablehead{\colhead{Star Name} & \colhead{Si~{\footnotesize III}} & \colhead{N~{\footnotesize V}} & 
\colhead{O~{\footnotesize I}} & \colhead{O~{\footnotesize I}]} \\[3mm]
\colhead{ } & \colhead{(1206~\AA)} & \colhead{(1240~\AA)} & 
\colhead{(1305~\AA)} & \colhead{(1355~\AA)}   \\[3mm]
\colhead{(1)} & \colhead{(2)} & \colhead{(3)} & \colhead{(4)} & \colhead{(5)}  } 
\startdata
 $\alpha$~Lep     &      0.26:    &    0.28   &  0.61:  & \nodata   \\[3mm]
 $\iota$~Car        &      0.37     &    0.36   &  1.0     &   0.23      \\[3mm]
 $\theta$~Sco     &      0.095:  &    0.052 &  0.36:  &  \nodata  \\[3mm]
 \enddata
\tablecomments{Cols.~2--5 are observed fluxes at Earth in $10^{-13}$ erg cm$^{-2}$ s$^{-1}$, without compensation for reddening.  Colons indicate uncertain values: due to weak feature, bright continuum, or spectral complexity (e.g., strong interstellar and/or circumstellar absorptions). N~{\footnotesize V} ~1240~\AA\ (Col.~3) is the sum of components 1238~\AA\ and 1242~\AA, and O~{\footnotesize I}~1305~\AA\ (Col.~4) is the sum of components 1304~\AA\ and 1306~\AA.   No uncertainties are cited for the integrated fluxes, because difficult-to-quantify systematic errors -- continuum placement and circumstellar effects -- dominate over the (negligible) photometric noise.}
\end{deluxetable}

\clearpage
\begin{figure}[ht]
\figurenum{4}
\vskip  0mm
\hskip  -5mm
\includegraphics[width=\linewidth]{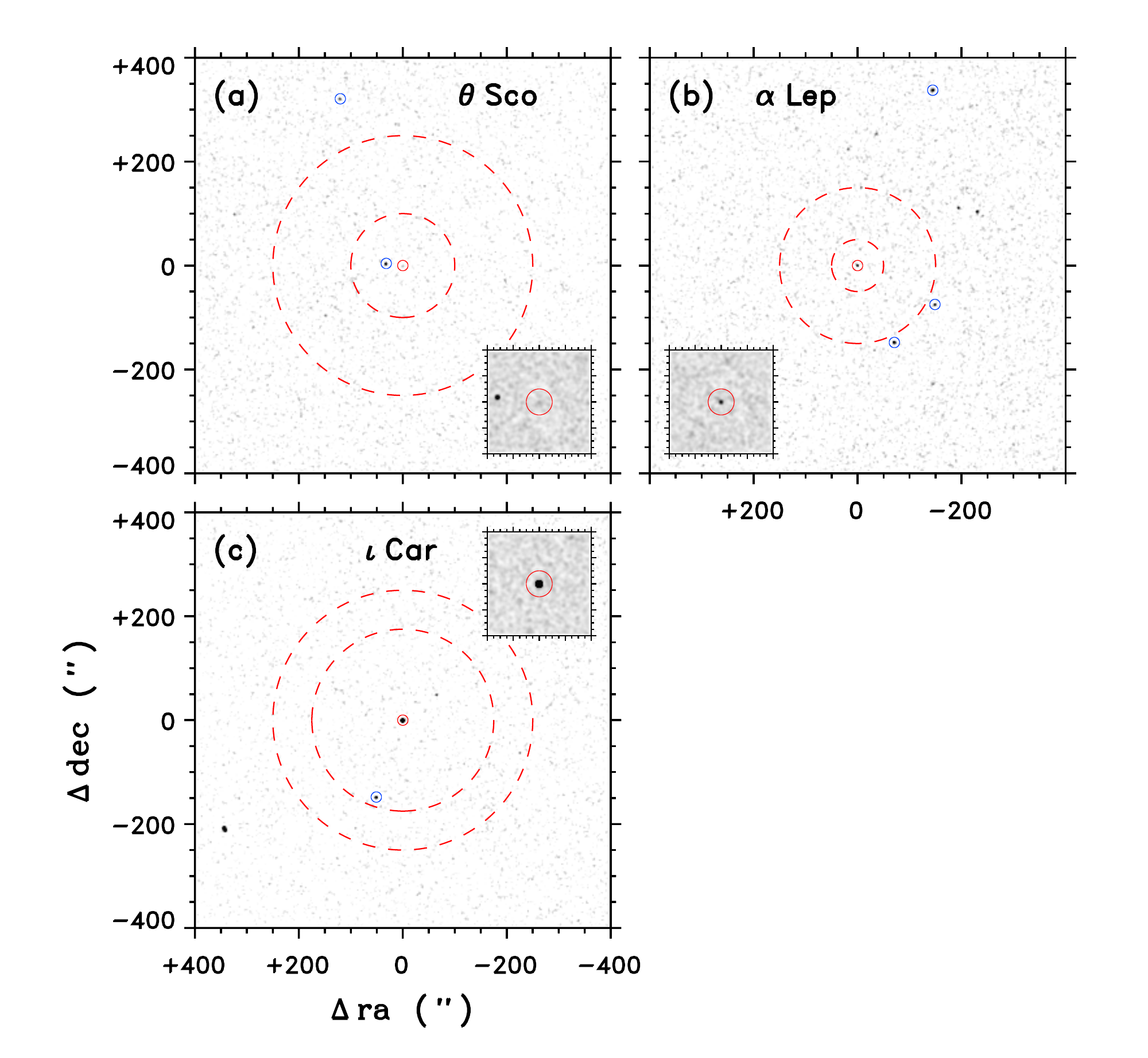} 
\vskip  0mm
\caption[]{\small
X-ray maps, binned in 0.5\arcsec\ pixels, of the fields around the three F supergiants in the {\em Chandra}\/ HRC-I program.  In each frame, a small red circle marks the position of the target; the central view is expanded in the inset panels (latter are 80\arcsec\ on a side).  In the main frames, larger dashed red circles indicate the inner and outer boundaries of the annulus in which the average background was determined.  The $\iota$~Car source had about 1000 net counts in 19~ks of exposure; the $\alpha$~Lep source about 40 counts, also in 19 ks; while $\theta$~Sco was very weak, with only 9 net counts in nearly 20~ks of exposure.  Blue-circled points are astrometric check objects: X-ray sources with catalogued optical (or near-IR) counterparts.
}
\end{figure}

\clearpage
\subsection{{\em Chandra}\/ HRC-I Imaging}

The three new F-supergiant targets all had been noted previously as X-ray sources:  $\iota$~Car from the {\em ROSAT}\/ All-Sky Survey (RASS) Faint Source Catalog (60 count ks$^{-1}$); $\theta$~Sco from the {\em ROSAT}\/ High Resolution Imager (HRI: 3 count ks$^{-1}$, but no detection in the RASS); and $\alpha$~Lep from the {\em XMM}\/ Slew Survey (although the cited soft-band [0.2--2~keV] intensity was considerably higher than expected given the absence of the target in the RASS).  The slim detections and inhomogeneous origins did not inspire confidence in the X-ray fluxes, or associated source positions; especially given that companions, possibly resolved, might provide the majority of the high-energy events.

To obtain secure X-ray detections, and positional confirmations, {\em Chandra}\,'s HRC-I has the advantages of excellent 1$\arcsec$ imaging (compared to, say, $\sim$10$\arcsec$ {\em XMM-Newton}\,); good low-energy response (important for soft stellar coronal sources); and freedom from ``optical-loading'' for visually bright, but possibly X-ray faint, stars.  {\em Chandra}\,'s other camera system, CCD-based ACIS, has poorer soft response (due to contamination build-up), and a ``red leak'' that hampers its use for optically bright objects like the F supergiants (all with $V<3$).  The poor -- essentially nonexistent -- energy resolution of HRC-I was not a concern, because this primarily was a detection experiment (and to assess whether any apparent source truly was coincident with the bright star).  In fact, these objects were not expected to be X-ray bright in the first place (see above), so an observation with an energy-resolving camera likely would not have collected a useful spectrum in the planned relatively short exposures.

The HRC-I pointings were carried out 2017 March ($\theta$~Sco), April ($\alpha$~Lep), and June ($\iota$~Car).  The approved exposures were  20~kiloseconds (ks) each, with 18.9--19.8~ks collected during the actual observations (accounting for dead time: see Table~2). 

Time-integrated X-ray event maps, binned in 0.5$\arcsec$ spatial pixels, are illustrated in Figure~4.  The predicted target location is at the center of each frame; N is up, E to the left; both axes in arcseconds.   A close-up of the central region of each field is shown in the insets.  There were clear detections of $\iota$~Car and $\alpha$~Lep, and a possible weak source at $\theta$~Sco.  An $r=1.5\arcsec$ detect cell (95\% encircled energy) was chosen as a compromise between maximizing the source counts while minimizing the uncertainty in the encircled energy correction, but also avoiding an excessive amount of background in the cell (see, e.g., Ayres 2004)\footnote{The 95\% encircled energy radius here is slightly smaller than the 1.6\arcsec\ used previously (e.g., A17).  The updated value is based on a re-assessment of the large collection of archival HRC-I exposures of AR~Lac (a hard source) and Procyon (soft source), including off-axis images of the former.  Concatenated event lists were measured (in concentric circles, accounting for background) to estimate 95\% encircled energy radii, on-axis as well as off-axis.  Positional uncertainties, utilizing the 95\% cell, as functions of the target offset and detection signal-to-noise ($\sqrt{N}$, where $N$ is the number of counts in the detect cell, and assuming negligible background) were deduced from Monte Carlo simulations, with shuffled event lists, of the source centroiding process.  Results can be summarized as follows.  For 95\% encircled energy, $r\sim 1.5 + 0.07\,{\rho}^{2.2}$~(\arcsec), where $\rho$ is the off-axis distance of the source in arcminutes.  The corresponding positional uncertainty ($\sqrt{\sigma_{\Delta{x}}^2\,+\,\sigma_{\Delta{y}}^2}$~) is, $(0.5\,+\,0.04\,{\rho}^{2.2})\,/\,\sqrt{N}$ (\arcsec).  For $\sigma_{\Delta{x}}$ and $\sigma_{\Delta{y}}$, independently (reported as a single value in Table~4), the uncertainties are 80\% of the previous relation.  No difference was found between the hard and soft sources as far as these quantities were concerned.  However, the off-bore-sight parameters were determined from event lists that were averaged over the four quadrants where the calibration exposures (of AR~Lac) were taken, so the specific asymmetry of an actual off-axis source was not accounted.  Thus, the encircled energy radii and positional uncertainties likely become less reliable, the further off-axis the source is located.}.  

An average cosmic background was determined in an annulus, between the inner and outer radii listed in Table~4, amounting to typically only 5~counts in the detect cell.  The measurements are summarized in Table~4, which also lists the offset of the detected central source with respect to the stellar coordinates, as validated by optical counterparts of X-ray check objects (marked in Fig.~4; best match also named in the Table).  Iota~Car was the strongest source, with more than 1000 net counts, as anticipated from the RASSFSC.  Alpha~Lep was second, although with only about 40 net counts, more than an order of magnitude fainter than expected from the reported {\em XMM-Newton}\/ Slew-Survey detection.  Both source positions were consistent with the optical coordinates of the bright stars, as validated by the astrometric check objects, so there was no overt evidence for resolved X-ray companions in either case.

Theta~Sco was faintest of the three, with only about 9 net counts, much weaker than anticipated from the reported HRI source, but nevertheless significant with respect to the background at a 3\,$\sigma$ level (0.13\% chance of false positive [one-sided Gaussian]: see Ayres 2004, his equation~4).  In fact, there is a second, stronger source (3.5 count ks$^{-1}$) about 30\arcsec\ East of $\theta$~Sco, which likely is the HRI detection previously classified as the F supergiant.  This source coincides with a {\em Gaia}\/ catalog entry, a relatively bright object with $G= 14.22$, but no proper motion or parallax reported in Data Release 1 (DR1).  For this reason, a second check star [see Table~4] was preferred, even though it was much further off the bore sight, because it had a full astrometric solution in DR1.  This second object is red, with $(B-V)\sim 0.9$; bright, with $G= 10.19$; and distant, at about 500~pc.  For that distance, the X-ray luminosity would be $\log{L_{\rm X}}\sim 30.1$ erg s$^{-1}$, assuming a hot ($10^7$~K) coronal source.  Given the apparent optical brightness, the second object likely is a late-type subgiant; and given the high X-ray luminosity, it probably is a short-period binary.  For the specific case of $\theta$~Sco, the target position was taken from the measured offset of the second check star.  This led to a slightly larger number of counts (14) versus the 13 collected when the centroid of the apparent $\theta$~Sco source ([$-$0.5\arcsec,+0.2\arcsec]) was used.  Under the latter scenario, the source significance still would be $3\,\sigma$, and the X-ray flux only slightly lower.  Because the $\theta$~Sco source was consistent with the stellar coordinates, under either scenario, there was no indication for a resolved X-ray companion.

\begin{deluxetable}{cclccc}
\rotate
\tabletypesize{\footnotesize}
\tablenum{4}
\tablecaption{X-ray Measurements from {\em Chandra}\/ HRC-I}
\tablecolumns{6}
\tablewidth{0pt}
\tablehead{\colhead{Object} & \colhead{$\rho$\,(\arcmin)} & \colhead{[$\Delta{x}$(\arcsec), $\Delta{y}$(\arcsec)]} & 
\colhead{Count Rate~(ks$^{-1}$)}  &  \colhead{$f_{\rm X}$}  &  \colhead{$\log{L_{\rm X}}$} \\
\colhead{(1)} & \colhead{(2)} & \colhead{(3)} & \colhead{(4)} & \colhead{(5)}  & \colhead{(6)}  
} 
\startdata
\cutinhead{ObsID\,17723: $b\sim 2.3{\times}10^{-2}$ count ks$^{-1}$ $({\arcsec})^{-2}$ (50\arcsec--200\arcsec)}
$\alpha$~Per      &  0.3 & [+0.7, +0.3] &  18.7${\pm}$0.9 & 2.2${\pm}$0.1& 29.80 \\ 
{\em Gaia}~441700724556170240~($G=\,12.22$)    &  4.2 &  [+0.8, +0.3]~(${\pm}0.1$) & 20.1${\pm}$1.0 & \nodata  & \nodata \\
\cutinhead{ObsID\,18912: $b\sim 3.7{\times}10^{-2}$ count ks$^{-1}$ $({\arcsec})^{-2}$ (50\arcsec--150\arcsec)}
$\alpha$~Lep      & 0.5   &   [+0.3, $-$0.1]  &  2.2${\pm}$0.4  &  0.18${\pm}$0.03 & 30.00 \\ 
2MASS~J05323339$-$1750351~($J=\,12.74$) & 2.8 & [+0.3, $-$0.1]~(${\pm}0.1$) & 2.6${\pm}$0.4 & \nodata  & \nodata \\
\cutinhead{ObsID\,18913: $b\sim 3.5{\times}10^{-2}$ count ks$^{-1}$ $({\arcsec})^{-2}$ (175\arcsec--250\arcsec) }
$\iota$~Car        & 0.5   &  [+0.1, $-$0.6] &   55${\pm}$2  &  5.0${\pm}$0.2 & 30.52\\ 
{\em Gaia}~5300299847292105088~($G=\,19.88$)    &  2.6 &  [$-$0.1, $-$0.6]~(${\pm}0.1$) & 4.7${\pm}$0.5 & \nodata  & \nodata \\
\cutinhead{ObsID\,18914: $b\sim 3.4{\times}10^{-2}$ count ks$^{-1}$ $({\arcsec})^{-2}$ (100\arcsec--250\arcsec)}
$\theta$~Sco     & 0.5  & ([$-$0.1, $-$0.4]) & 0.5$^{+0.3}_{-0.2}$ &  0.06$^{+0.03}_{-0.02}$ &  27.78\\ 
{\em Gaia}~5955859633587752192~($G=\,10.19$)    &  5.7 &  [$-$0.1, $-$0.4]~(${\pm}0.4$) & 2.7${\pm}$0.5 & \nodata & \nodata \\
\enddata
\vskip -3mm
\tablecomments{In each observation ({\em Chandra}\/ ObsID) heading, $b$ is the average background (rate) measured in an annulus (parenthetical values) centered on the target.  Col.~2 is the separation of the object from the image center.  Col.~3 is the offset of the X-ray centroid from the optical coordinates of the object.  For $\theta$~Sco, only, the offset was taken from the check star.  Col.~5 is the X-ray flux at Earth in $10^{-13}$ erg cm$^{-2}$ s$^{-1}$.  Col.~6 is the X-ray luminosity, in erg s$^{-1}$.}
\end{deluxetable}

Extinction-dependent Energy Conversion Factors (ECF) were applied to the source net count rates (CR), adjusted for the 95\% encircled energy fraction, to obtain energy fluxes at Earth.  The ECF values ($7.9{\times}10^{-12}$ erg cm$^{-2}$ count$^{-1}$ for $E(B-V)=+0.02$ [$\alpha$~Lep] to 11.0${\times}10^{-12}$ for $E(B-V)=+0.15$ [maximum-reddened $\alpha$~Per]) were derived using the {\em Chandra}\/ WebPIMMS tool (Cycle~18 version)\footnote{See: http://cxc.harvard.edu/toolkit/pimms.jsp} for a $T= 10^{7}$~K, solar abundance APEC plasma model and energy range 0.2--2~keV for the unabsorbed flux.  Although the coronal temperatures of the three new F supergiants could not be determined (as mentioned earlier, HRC-I lacks spectral discrimination), related object Canopus was known to be hot ($\sim 10^7$~K: from a  previous {\em Chandra}\/ HETGS spectrum [Brown et al.\ 2003]), as well as $\alpha$~Per from its {\em ROSAT}\/ PSPC hardness ratio; both of which motivated the choice of the baseline ECF temperature.  Nevertheless, the HRC-I response is relatively independent of temperature from $10^7$~K down to a few million K; uncertainty in the hydrogen column generally amounts to a larger factor.  Even so, a 0.5~dex uncertainty in $N_{\rm H}$ translates to an only 0.1~dex uncertainty in the ECF, for $T\sim 10^{7}$~K and $N_{\rm H}$ near $3{\times}10^{20}$ cm$^{-3}$ ($E(B-V)\sim 0.05$). 

\clearpage
\begin{figure}[ht]
\figurenum{5a}
\vskip  0mm
\hskip  -5mm
\includegraphics[width=\linewidth]{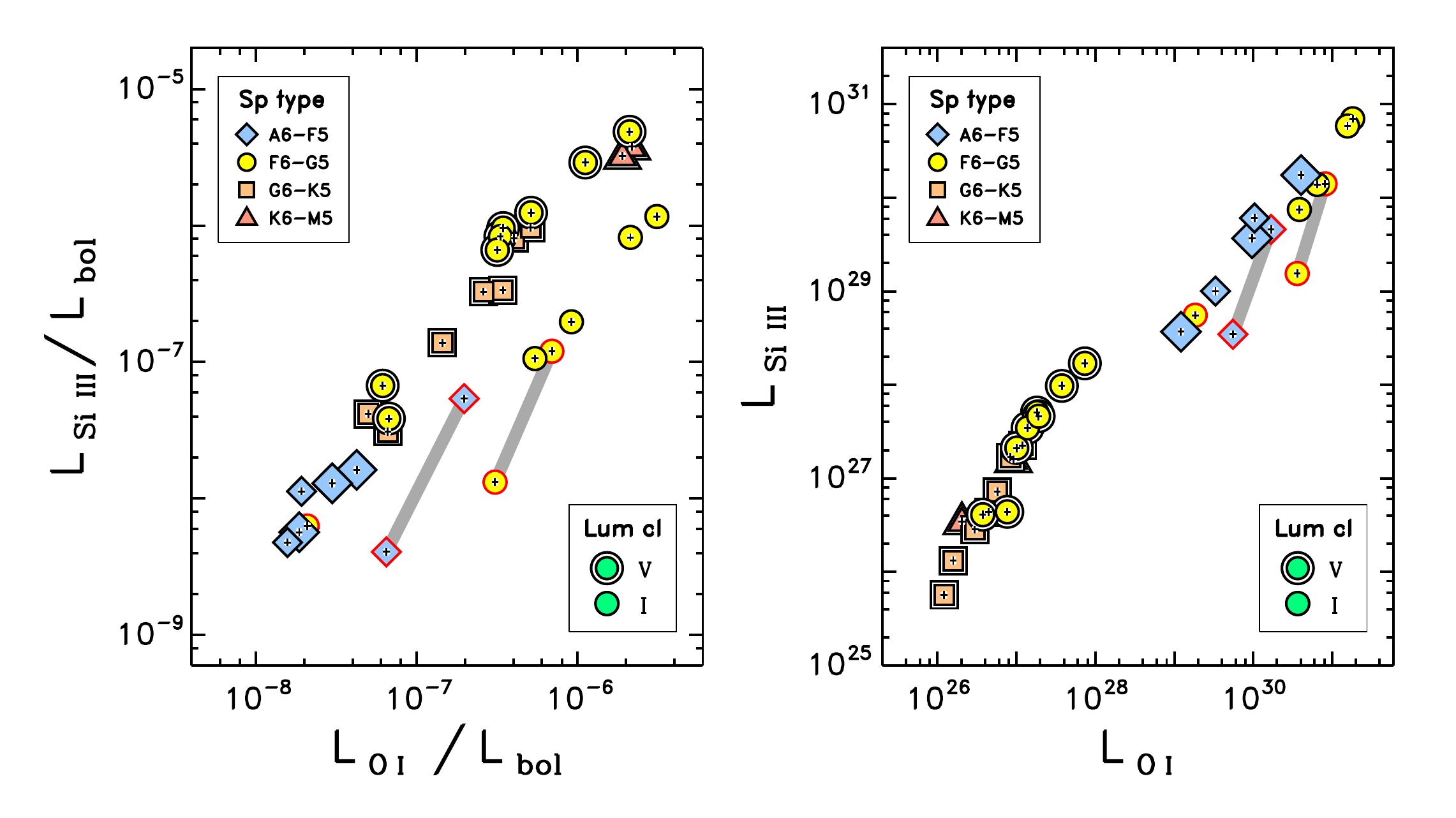} 
\vskip  0mm
\caption[]{\small
{\em Left:}\/ $L_{\rm Si\,III}/L_{\rm bol}$ versus $L_{\rm O\,I}/L_{\rm bol}$ for F and G supergiants and G--M dwarfs, according to the two legends.  {\em Right:}\/  Alternative {Si}\,{\scriptsize III}/{O}\,{\scriptsize I} flux-flux diagram, expressed in absolute luminosities.  In this case,``Si \,{\scriptsize III}'' refers to the resonance line at 1206~\AA, while ``{O}\,{\scriptsize I}'' refers to the two longward components of the resonance triplet at 1305~\AA.
}
\end{figure}

\section{ANALYSIS}

The next several figures are ``flux-flux'' diagrams, which compare one emission species against another, usually spanning widely separated formation temperatures.  Such diagnostic diagrams provide insight concerning how different ``layers'' of the outer atmosphere are energetically connected.  In what follows, the left hand panels display line fluxes normalized to bolometric fluxes (equivalently, $L/L_{\rm bol}$), to allow a fairer comparison among stars of different sizes and distances.  For balance, absolute luminosities are shown in the right hand panels.  Spectral types and luminosity classes are encoded in the figures by symbols and colors according to the two legends.  The three subjects of the present study are the larger blue diamonds (typically in the lower left side of the left panel); the two previous objects of interest -- $\alpha$~Per and Canopus -- are the smaller blue diamonds.  Symbols outlined in red are Cepheid variables.  Diagonal shaded bars connect the FUV (or X-ray) low and high states for the two Cepheids that have phase-resolved high-energy measurements (as noted earlier).

The bolometrically normalized fluxes indicate how efficiently the star is able to divert its total energy flow -- mainly visible light -- into the exotic high-energy emissions, a process thought to be mediated by surface magnetic fields.  Young, active G dwarfs are rather good at this, with perhaps 0.1\% of their total luminosity ending up at high energies; whereas inactive stars like the Sun are rather poor, barely managing to convert a ten-millionth ($10^{-7}$) of their total energy flow into coronal emissions. 

The left hand side of Figure~5a depicts $L_{\rm Si\,III}/L_{\rm bol}$ versus $L_{\rm O\,I}/L_{\rm bol}$ for selected G--M dwarfs and F and G supergiants.  The heritages of the flux measurements can be found in A17, for the bulk of the objects not described in the present study.  The FUV fluxes were corrected for reddening, prior to assembling the diagrams, according to the wavelength-dependent average galactic extinction formula of Fitzpatrick \& Massa (2007), utilizing the color excesses reported here or in A17.  \ion{Si}{3} 1206~\AA\ is a hot sub-coronal diagnostic, more connected to the presumably exclusively magnetically heated corona; whereas \ion{O}{1} 1305~\AA\ is representative of the cooler chromospheric layers, which are thought to be energized mainly by the underlying photosphere, partly by acoustic shocks as well as interactions between small-scale magnetic structures buffeted by photospheric flows (as, for example, in the solar supergranulation network).

This is a remarkable comparison, because all the F supergiants, including low-amplitude Cepheid Polaris ($\alpha$~UMi: F8~Ib), cluster at the terminus of the dwarf star power law, on the lower side, rather than following the displaced (to the right) trend of the G supergiants, and possibly also the FUV high states of the two phase-resolved Cepheids.  The FUV low states of the Cepheids, in this comparison, shift to the lower left, perhaps pointing to the true extension of the G-supergiant trend toward lower activity.  Previously, A17 suggested that $\alpha$~Per and Canopus might sit at the extreme low end of a leftward extension of the G-supergiant track, an association that still plausibly could be made (if one ignores the two Cepheid FUV low states).  However, the new F supergiants do not fill in the putative extrapolation of the G-supergiant sequence, but rather congregate together, as if they instead were following a lower extension of the dwarf-star power law; now an equally plausible, though surprising, possibility (so odd, it was not considered seriously in A17).

The exaggerated chromospheric oxygen emissions of the G supergiants, at similar sub-coronal activity levels to the dwarf stars, could come about from a variety of factors.  An important one is density.  The cooler supergiants likely have significantly lower chromospheric densities, so Bowen-fluorescence pumping of the oxygen resonance lines by \ion{H}{1} Ly$\beta$ (e.g., Carlsson \& Judge 1993) could be strongly enhanced.  A likely more important factor is the atmospheric extent.  A G-supergiant chromosphere is proportionately much thicker (relative to the stellar radius) than that of a dwarf star (a scale-height effect: see Ayres et al.\ 2003), which allows the optically thick oxygen emissions to be produced throughout a larger volume and scatter to the surface, boosting the overall intensity.  The F supergiants seem to side with the dwarfs in this respect, which comes back to the carbon multiplet absorption/emission dichotomy mentioned earlier:  the G-supergiant chromospheres are thick enough to allow the atomic carbon lines to go into emission; while the F supergiant chromospheres apparently are too thin, so the two carbon multiplets arise in the photosphere, in absorption, instead.

\begin{figure}[ht]
\figurenum{5b}
\vskip  0mm
\hskip  -5mm
\includegraphics[width=\linewidth]{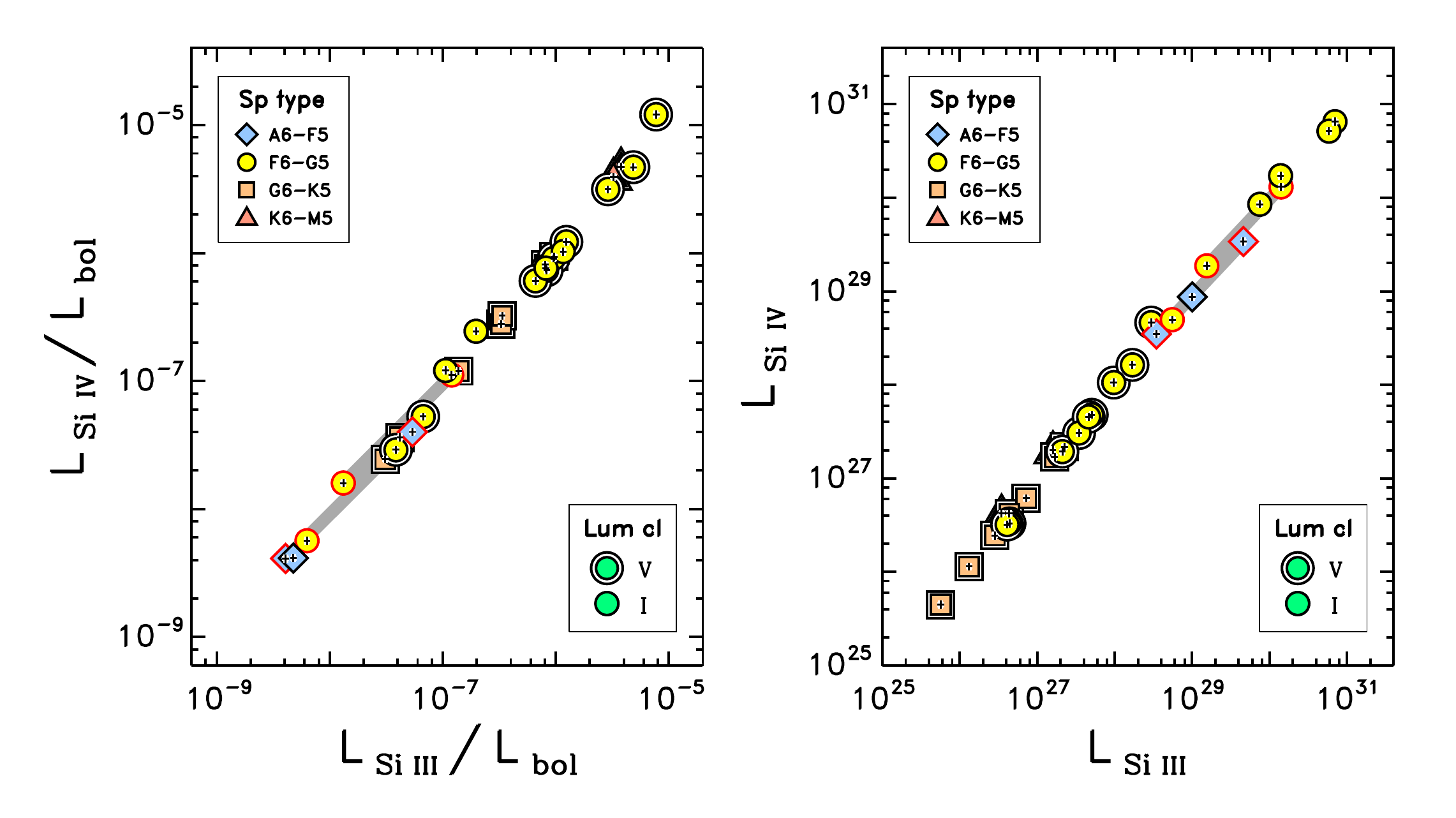} 
\vskip  0mm
\caption[]{\small
Same as Fig.~5a for Si~{\scriptsize IV} 1400~\AA\ versus Si~{\scriptsize III} 1206~\AA.
}
\end{figure}

However, when projected onto the absolute luminosity plane (right panel), the F supergiants now seem to overlap the lower end of the G-type sequence, despite the large apparent separation in the bolometrically normalized comparison on the left.  The warm F supergiants shift upward significantly, toward the right, in the absolute luminosity diagram on a 1:1 track, and thus fall away, on the low side, from the steeper \ion{Si}{3}/\ion{O}{1} power law traced by the dwarfs.  That the F supergiants and their G counterparts seem to inhabit the same region in the absolute luminosity diagram might be coincidental, or could be saying something more fundamental concerning their outer atmospheres.  On the one hand, these comparisons imply that the absolute sub-coronal and chromospheric luminosities of the F and G supergiants are similar, despite the large differences in the photospheric properties of the two classes.  On the other hand, the bolometrically normalized fluxes push the F supergiants down to the lower side of the dwarf star track, which indicates that the efficiency of converting their total energy flow into high-energy emissions is as low, or lower, than the lowest activity late-type dwarfs (like the Sun, for example).  Meanwhile, the G supergiants sport order of magnitude or higher conversion efficiencies on both the sub-coronal and chromospheric fronts.  

Incidentally, the similarity in absolute chromospheric luminosities of the F and G supergiants, and clean separation from the dwarf sequence, demonstrates that the F-supergiant emission is intrinsic to the luminous stars, rather than, say, representing a line-free supergiant energy distribution with a superposed emission-line spectrum from an unresolved hyperactive dwarf companion.  This is a reason enough to include the absolute-$L$ diagrams together with the normalized $L/L_{\rm bol}$ versions.

Figure~5b depicts sub-coronal \ion{Si}{4} 1400~\AA\ versus sub-coronal \ion{Si}{3}~1206~\AA.  Note that none of the new F supergiants were detected in \ion{Si}{4}, nor was Canopus previously, so they all are absent from the figure.  The bolometrically normalized luminosities, in the left panel, show the two diagnostics locked in a 1:1 embrace, for dwarfs and supergiants alike, even despite the possible presence of circumstellar absorptions in the \ion{Si}{3} features of the more luminous objects.   The connected FUV low/high states of the (two) Cepheids fall nearly exactly on the same trend.  When translated to absolute fluxes (right panel), which simply slides all the supergiants up the 1:1 line, the same close correspondence is maintained.  This shows that \ion{Si}{3} -- easily detected in all the study and reference stars above the reduced continuum emission at 1200~\AA\ -- can be used as a proxy for the higher-temperature \ion{Si}{4} doublet, which often is swamped at 1400~\AA\ by continuum light in the warm supergiants.

\begin{figure}[ht]
\figurenum{5c}
\vskip  0mm
\hskip  -5mm
\includegraphics[width=\linewidth]{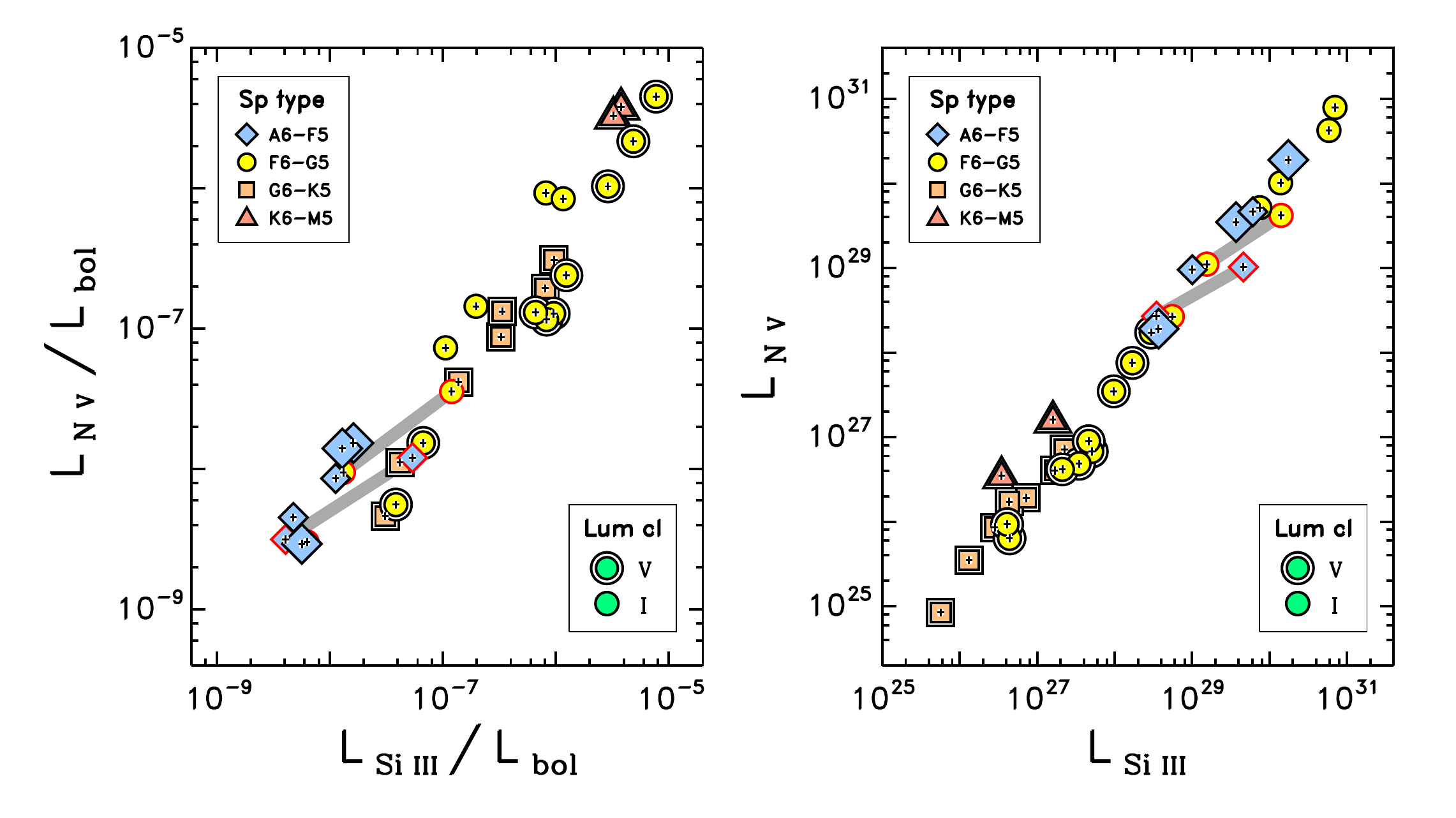} 
\vskip  0mm
\caption[]{\small
Same as Fig.~5a for N~{\scriptsize V} 1240~\AA\ versus Si~{\scriptsize III} 1206~\AA.
}
\end{figure}

Figure~5c compares \ion{N}{5}~1240~\AA, highest temperature sub-coronal species described here, to \ion{Si}{3}~1206~\AA.  In the left hand panel (bolometrically normalized luminosities), the normal F supergiants, and the Cepheid FUV low states, again cluster together, alongside the bottom of the dwarf track, displaced to the left, in a roughly 1:1 power law.  The G supergiants also fall slightly to the left of the dwarf track, but higher up.  Given that the \ion{Si}{4}/\ion{Si}{3} comparison was essentially 1:1, the drift to the left of the evolved stars (which means higher nitrogen emission at a given silicon emission) likely is an abundance effect.  In fact, such stars are known to be nitrogen rich owing to post-MS dredge-up episodes (Luck \& Lambert 1985).  Given the potential abundance bias, \ion{N}{5} is less favored as a sub-coronal tracer than \ion{Si}{3}, which is minimally affected by evolution-induced compositional changes.

\begin{figure}[ht]
\figurenum{5d}
\vskip  0mm
\hskip  -5mm
\includegraphics[width=\linewidth]{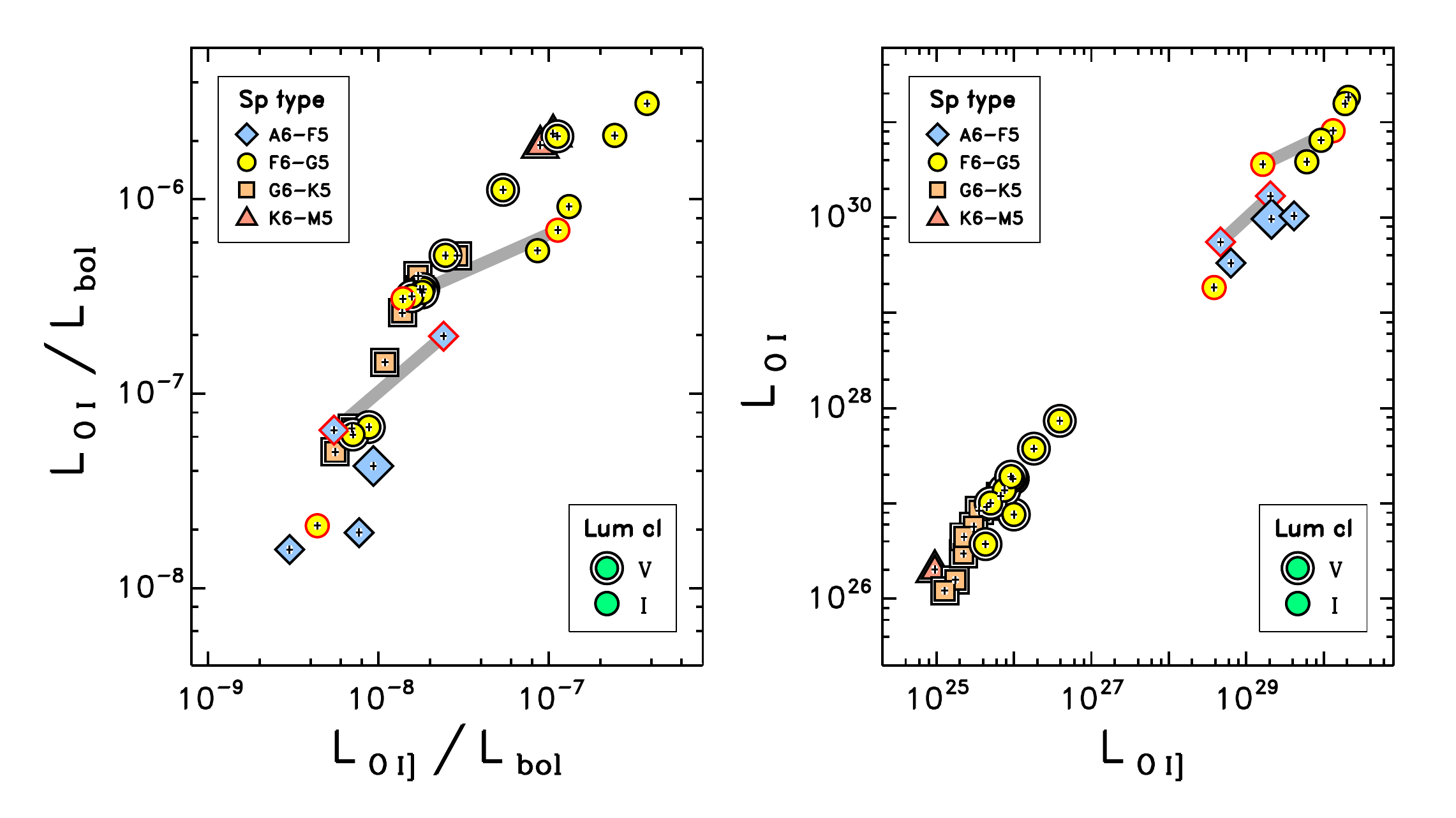} 
\vskip  0mm
\caption[]{\small
Same as Fig.~5a for O~{\scriptsize I} 1305~\AA\ versus semi-permitted O~{\scriptsize I}] 1355~\AA.
}
\end{figure}

Figure~5d illustrates the \ion{O}{1} 1304~\AA\ + 1306~\AA\ resonance lines versus semi-permitted \ion{O}{1}]~1355~\AA, both presumably formed under chromospheric conditions.  This diagram is a link to A17, where the prominent \ion{O}{1}]~1355~\AA\ features of the supergiants (and dwarfs) were utilized as a chromospheric proxy, preferred then over the oxygen resonance lines (1305~\AA) due to high opacity of the latter and possible circumstellar absorptions.  Unfortunately, the trend of conspicuous semi-permitted oxygen emission seen previously did not extend to the three new F supergiants: \ion{O}{1}] was detected in only one, $\iota$~Car.   Nevertheless, the comparison is instructive:  the F supergiants and Cepheid Polaris again cluster at the bottom, possibly slightly to the right, of the G-dwarf track; while the G supergiants, and Cepheid FUV high states, are displaced upward and to the right, showing excess semi-permitted oxygen emission.  This could be a signature of very low density conditions in which the \ion{O}{1}]~1355~\AA\ emission is dominated by recombination.  It also suggests that \ion{O}{1}]~1355~\AA\ perhaps was not the best choice for a ``chromospheric'' proxy in the previous study, since there apparently is an additional bias for the cooler supergiants.

\begin{figure}[ht]
\figurenum{5e}
\vskip  0mm
\hskip  -5mm
\includegraphics[width=\linewidth]{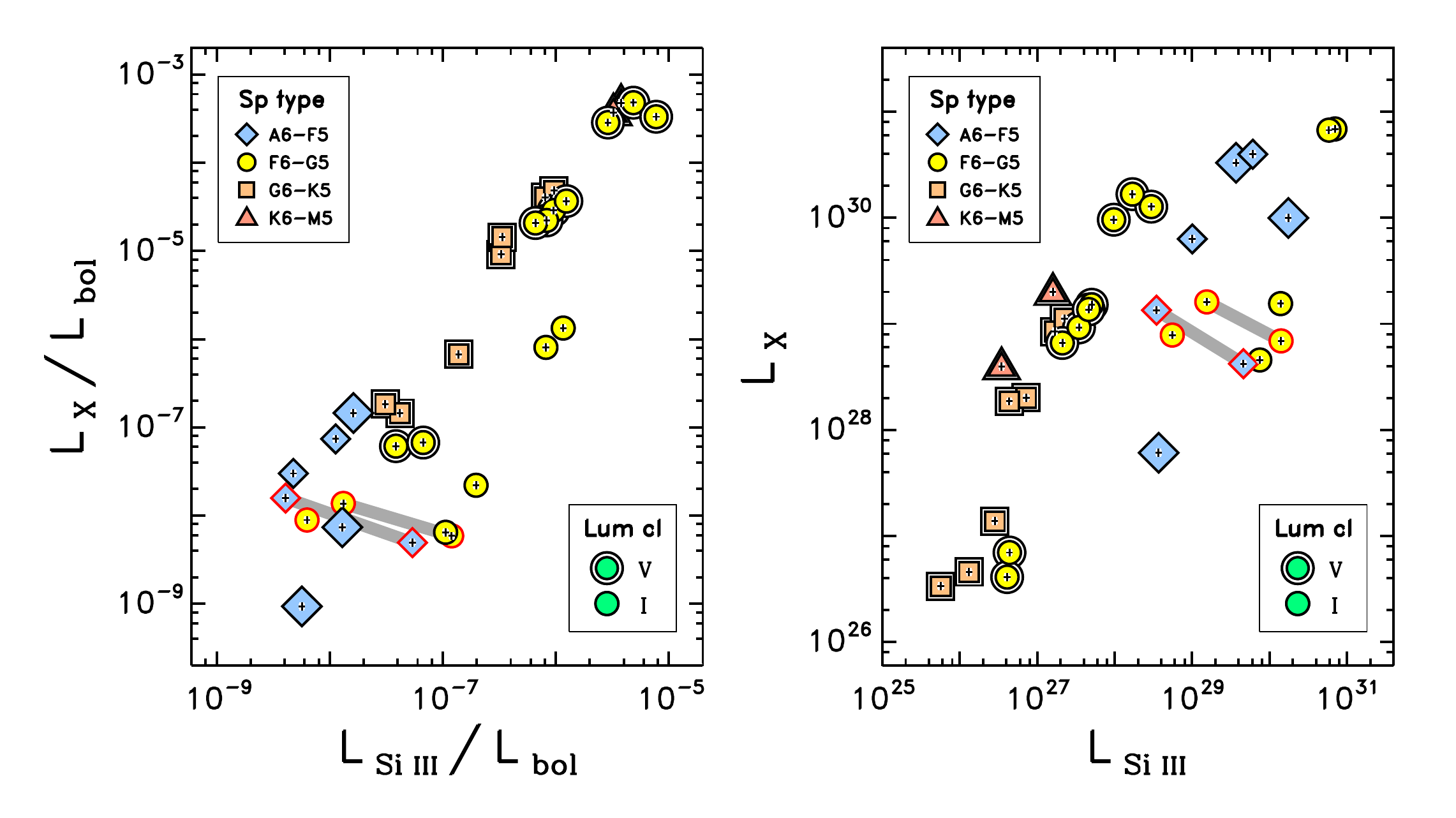} 
\vskip  0mm
\caption[]{\small
Same as Fig.~5a for X-rays (0.2--2~keV) versus Si~{\scriptsize III} 1206~\AA.}
\end{figure}
 
Figure~5e depicts X-rays (0.2--2~keV) versus \ion{Si}{3} 1206~\AA.  This was the pivotal comparison in A17 (with \ion{Si}{3} here replacing the \ion{Si}{4} used there), which -- absent the new targets -- suggested that $\alpha$~Per and Canopus were unusual in their coronal/sub-coronal behavior, at least compared to the several examples of G supergiants, of various activity levels; in the sense that the latter are shifted well to the right of the dwarf sequence, whereas the former two F supergiants fell somewhat to the left.  Now, the additional three new F supergiants appear to occupy more-or-less the same region as the previously ``anomalous'' $\alpha$~Per and Canopus, thus reinforcing the idea that the F supergiant class itself is discrepant in X-rays relative to the cooler members of the evolved intermediate-mass stars.  

As noted in A17, it is curious that the Cepheid X-ray high states follow the flux-flux behavior of the F-supergiants, whereas the FUV high states align more closely with the G supergiants.  This is all the more remarkable because the Cepheid case seems to be strongly linked to their atmospheric pulsations, given the temporally sharp X-ray and FUV enhancements, although with substantially different phasing (Engle et al.\ 2017).  The more normal F and G supergiants do not ostensibly pulsate nor are obviously time-variable (although note, as mentioned earlier, low-amplitude Cepheid Polaris [isolated red-circled yellow dot] sits with the F supergiants).  Despite the relatively organized structure of the left hand panel, the right panel (absolute luminosities) shows a more chaotic blob of points to the right of the tighter correlation exhibited by the dwarf stars.  Notably, the three new F supergiants still have X-ray luminosities that are similar to, or much less than, active dwarf stars of comparably young age.  There is no reason to suppose that the F supergiants could not have $L_{\rm X}$ an order of magnitude, or more, higher than the most X-ray luminous young G dwarfs, but apparently this is not the case (at least in the small sample to date).  Again, perhaps a coincidence, but also possibly an important piece of a larger puzzle.

\begin{figure}[ht]
\figurenum{5f}
\vskip  0mm
\hskip  -5mm
\includegraphics[width=\linewidth]{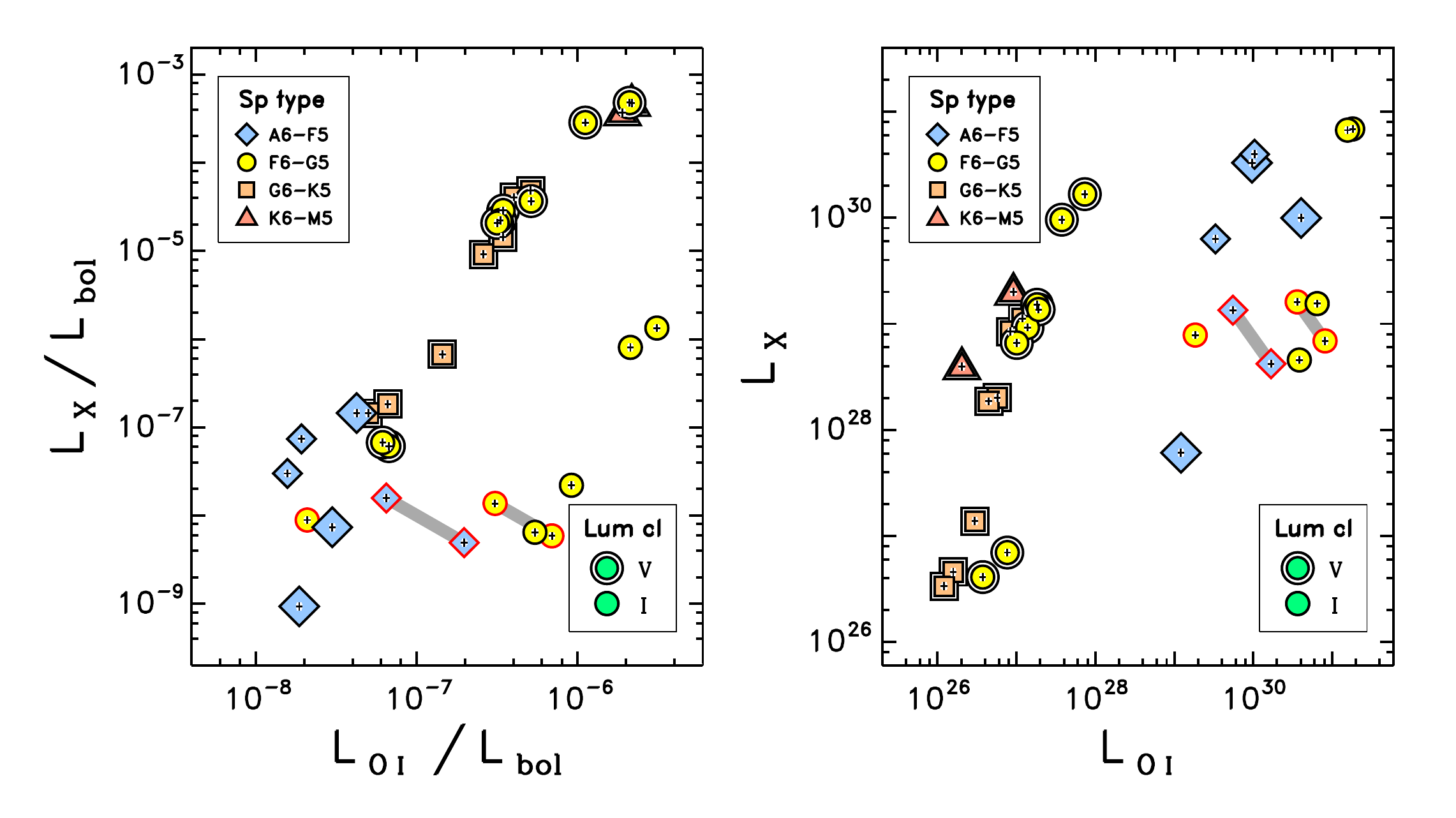} 
\vskip  0mm
\caption[]{\small
Same as Fig.~5a for X-rays (0.2--2~keV) versus O~{\scriptsize I} 1305~\AA.}
\end{figure}

The final diagram, Figure~5f, compares coronal X-rays to the chromospheric oxygen emission, the two diagnostics of this study furthest separated in formation temperature.  The bolometrically normalized luminosities in the left hand panel show a now familiar pattern.  The dwarf stars follow a steep power law (index $\sim$3) extending across much of the diagram; while the F supergiants, and Cepheid Polaris, appear to lie on an extension of the same track to even lower levels.  Meanwhile, the G supergiants fall on a parallel sequence displaced significantly to the right, toward higher chromospheric oxygen emission at similar coronal intensities.  Again, the two Cepheids with phase-resolved X-rays and FUV emissions seem to point alternatively to, and partially bridge, the F- and G-supergiant domains.   

The right hand panel, for the absolute luminosities, indicates that the supergiants, as a group, have much more chromospheric emission than dwarf stars, not surprising given the enormously larger surface areas of the former (one reason to favor the bolometric normalization); but, as noted before, the coronal X-ray luminosities span a similar range as the moderate to highly active dwarfs (the reason why binarity often has been invoked when an evolved object exhibits unusual coronal properties, as in the original study of $\alpha$~Per).

\section{DISCUSSION}

The new observations paint a compelling picture that the early-F supergiants, as a class, behave quite differently than their cooler G-type cousins only somewhat further along the 5--10~$M_{\odot}$ evolutionary tracks.  In fact, the F supergiants seem, surprisingly, to follow an extension of the X-ray/FUV flux-flux relationships defined by cool dwarfs, in spite of the vast gulf between their physical properties.  This, and other characteristics of their FUV spectra (weak Ly$\alpha$ emissions, narrow \ion{O}{1} resonance lines, and carbon multiplet absorption), suggest that the F supergiants have thin chromospheres, at least compared with the G-type supergiants.   For the latter, thicker chromospheres might partially smother high-energy emissions from hot coronal structures (magnetic ``loops'' in the solar context) embedded within the extended outer atmospheres, while the lower opacity in the FUV might allow any associated sub-coronal emissions at the longer wavelengths to escape relatively unscathed; thus resulting in the apparent G-supergiant ``X-ray deficiency'' (as noted observationally by Ayres et al.\ [2005]; see also Ayres et al.\ [2003] for a discussion of the ``buried corona hypothesis'').  

The thinness of the F-supergiant chromospheres could be due to weakening of what one might call the ``ionization valve'' effect, which normally allows height-extensive warm (6000--8000~K) temperature inversions to develop above the cool photospheres of late-type stars (Ayres 1979).  The mechanism works as follows.  

The upper photospheric temperatures of a cool star fall off monotonically toward higher altitudes in radiative equilibrium, creating a baseline, low ionization state of the outer atmosphere ($n_{\rm e}/n_{\rm H}\sim 10^{-4}$, dictated by the ``easily ionized'' metals Fe, Mg, and Si, for $T\lesssim 5000$~K).  The atmosphere can remain in balance at low temperatures, $< 5000$~K, because the (dominant) H$^{-}$ radiative heating and cooling (per gram) both depend on the electron density, $n_{\rm e}$, while the radiative heating itself falls outward as the gas becomes transparent, and the photons can escape into the dark sky away from the star. 

However, if there is extra, non-radiative, energy deposition at the top of the photosphere, and at higher altitudes, the gas cannot rid itself of the additional heat in its low ionization state, for the following reason.  The radiative cooling per gram of material is proportional to the electron density through H$^{-}$ associative attachment as well as collisional excitation of strong resonance lines, at least those that are not too optically thick at the base of the chromosphere.  The overall cooling is proportional to the absolute hydrogen density $n_{\rm H}$, thanks to the key role of the electron density, $n_{\rm e}\sim 10^{-4}\,n_{\rm H}$.  Both of these tightly coupled densities fall off rapidly with increasing height in a hydrostatic atmosphere for $T< 5000$~K.  Consequently, the temperatures must rise until the enhanced ionization boosts the electron density enough to bring the radiative cooling back into balance with the extra heating.  

Increasing ionization can keep up with a more-or-less height-independent heating (per gram), in the face of the rapid outward decline of $n_{\rm H}$, over many pressure scale heights, because when hydrogen begins to be stripped ($T\gtrsim 6000$~K), there are lots of electrons available (i.e., the ionization fraction can increase {\em four orders of magnitude}\/ from the low temperature limit of $10^{-4}$ at 5000~K, contributed by the metals, up to $\sim 1$ near 8000~K, where hydrogen is fully ionized).

Once the hydrogen ionization is almost complete, however, the gas no longer can respond to the mechanical heating by slowly increasing the temperature outward, but rather a catastrophic thermal instability must ensue, imposing an abrupt ``transition zone'' temperature rise at the top of the thick, nearly isothermal (6000--8000~K), chromosphere.  

The ionization valve mechanism cannot act as effectively in the F supergiants, or hotter stars in general, because there is less of a temperature contrast between the warm outer photosphere and the maximum hydrogen ionization temperature ($\sim 8000$~K, as mentioned above, or even lower under low-pressure conditions).  Thus, chromospheres on such stars could be very thin, in contrast to G supergiants only a couple of thousand degrees cooler.  (The ionization valve also operates effectively in cool dwarfs, like the Sun, where proportionately thinner chromospheres result none-the-less, but because of a scale-height effect due to high gravity: see, e.g., Ayres et al.\ 2003.)

In light of the new findings, perhaps the original ``Curious Case of the Alpha Persei Corona:\  A Dwarf in Supergiant's Clothing?'' (Ayres 2011), inspired by the (now out-of-favor) companion hypothesis, should be rephrased:  ``A Supergiant in Dwarf's Clothing?'' given that the F-supergiant outer atmospheric ``attire'' might mimic in some important respects the thin -- threadbare if you will -- aspects of a dwarf star.

\clearpage
\acknowledgments
This work was supported by grant GO6-17005X from the Smithsonian Astrophysical Observatory, based on observations from the {\em Chandra}\/ X-ray Observatory, collected and processed at the {\em Chandra}\/ X-ray Center, operated by SAO under contract to NASA; and grants GO-14484 ($\alpha$~Per) and GO-12278 (Advanced Spectral Library [ASTRAL]: Cool Stars) from the Space Telescope Science Institute, based on observations from {\em Hubble Space Telescope}\/ collected at STScI, operated by the Associated Universities for Research in Astronomy, also under contract to NASA.  This study also made use of public databases hosted by {SIMBAD}, maintained by {CDS}, Strasbourg, France, especially VizieR catalogue access to the {\em Gaia}\/ space astrometry mission Data Release 1; the Mikulski Archive for Space Telescopes at STScI in Baltimore, Maryland; and the High Energy Astrophysics Science and Research Center at the NASA Goddard Space Flight Center, in Greenbelt, Maryland.      


\end{document}